\documentclass{article}
\usepackage{hfsprocl}

\usepackage{axodraw}
\usepackage{rotating}
\usepackage{array}

\makeatletter
\def\section{\@startsection {section}{1}{\z@}{-3.5ex plus -1ex minus 
    -.2ex}{2.3ex plus .2ex}{\boldmath\bf }}
\makeatother

\renewenvironment{thebibliography}[1]
	{\begin{list}{\arabic{enumi}.}
	{\usecounter{enumi}\setlength{\parsep}{0pt}
	 \setlength{\itemsep}{0pt plus 5pt} 
         \settowidth
	{\labelwidth}{#1.}\sloppy\frenchspacing}}{\end{list}}

\newcommand{\MSbar}{\overline{\mbox{\small MS}}}
\newcommand{\aalpha}{\left(\frac{\alpha_s(m_b)}{\alpha_s(a^{-1})}\right)}

\def\dotprod{\mathord{\cdot}}
\def\vec#1{\mathbf{#1}}
\def\bra#1{\left\langle #1\right|}
\def\ket#1{\left| #1\right\rangle}
\newdimen\unit
\def\point#1 #2 #3{\vbox to0pt{\kern-#2\unit
  \hbox{\kern#1\unit$#3$}\vss}
 \nointerlineskip}
\def\btorho{\bar B^0\to\rho^+ l^- \bar\nu_l}
\def\btopi{\bar B^0\to\pi^+ l^- \bar\nu_l}
\def\btokstargamma{\bar B\to K^* \gamma}
\def\vub{|V_{ub}|}

\def\qsqmax{q^2_{\mathrm{max}}}
\def\w{\omega}

\def\ts{\vrule height2.5ex depth0pt width0pt}
\def\bs{\vrule height0pt depth4ex width0pt}

\newcommand{\PBS}[1]{\let\temp=\\#1\let\\=\temp}% latex companion p108
\def\etal{et al.}
\newcommand{\fd}{f_D}
\newcommand{\fds}{f_{D_s}}
\newcommand{\fb}{f_B}
\newcommand{\fbd}{f_{B_d}}
\newcommand{\fbs}{f_{B_s}}
\newcommand{\bb}{B_B}

\newcommand{\rgbb}{\hat{B}_B}
\newcommand{\rgbbd}{\hat{B}_{B_d}}
\newcommand{\rgbbs}{\hat{B}_{B_s}}

\def\kev{\,\mathrm{ke\kern-0.1em V}}
\def\mev{\,\mathrm{Me\kern-0.1em V}}
\def\gev{\,\mathrm{Ge\kern-0.1em V}}
\def\tev{\,\mathrm{Te\kern-0.1em V}}
\def\gtrsim{\mathrel{\lower .7ex\hbox{$\buildrel\textstyle>\over\sim$}}}
\def\lesssim{\mathrel{\lower .7ex\hbox{$\buildrel\textstyle<\over\sim$}}}
\newcommand{\csw}{c_\mathrm{sw}}
\newcommand{\kcrit}{\kappa_\mathrm{crit}}
\newcommand{\err}[2]{%
{{\renewcommand{\arraystretch}{0.4}%
\ensuremath{\mathop{\raisebox{0.1\height}{\scriptsize
$\begin{array}{@{}c@{}}+\\-\end{array}$}}%
\raisebox{0.1\height}{\scriptsize
$\begin{array}{@{}r@{}}#1\\#2\end{array}$}}}}%
}

\let\errp\errparen

\makeatletter
\def\slash#1{{\mathpalette\c@ncel{#1}}} % TeXbook, bottom of p360
\makeatother
\newcommand{\Dslash}{\slash D}%{\rlap{D}\hbox{\hspace{2pt}/}}
\newcommand{\lqcd}{\Lambda_{\mathrm{QCD}}}
\newcommand{\eqref}[1]{Eq.~(\ref{eq:#1})}
\newcommand{\secref}[1]{Section~\ref{sec:#1}}
\newcommand{\subsecref}[1]{Section~\ref{subsec:#1}}
\newcommand{\figref}[1]{Figure~\ref{fig:#1}}
\newcommand{\tabref}[1]{Table~\ref{tab:#1}}
\newcommand{\be}{\begin{equation}}
\newcommand{\ee}{\end{equation}}
\newcommand{\beqn}{\begin{eqnarray}}
\newcommand{\eeqn}{\end{eqnarray}}
\newcommand{\<}{\langle}
\renewcommand{\>}{\rangle}
\newcommand{\nn}{\nonumber}

\begin{document}
% start of shep titlepage
\begin{flushright}SHEP 97/20\\
hep-lat/9710057
\end{flushright}
\vspace{3em}

\begin{center}
{\Large\bfseries Heavy Quark Physics From Lattice QCD}\\[2em]
J.M.~Flynn and C.T.~Sachrajda\\[1em]
Department of Physics and Astronomy, University of Southampton\\
Southampton SO17 1BJ, UK
\end{center}
\vfill

\begin{center}\textbf{Abstract}\end{center}
\begin{quote}
We review the application of lattice QCD to the
phenomenology of $b$- and $c$-quarks. After a short discussion of the
lattice techniques used to evaluate hadronic matrix elements and the
corresponding systematic uncertainties, we summarise results for
leptonic decay constants, $B$--$\bar B$ mixing, semileptonic and rare
radiative decays. A discussion of the determination of heavy quark
effective theory parameters is followed by an explanation of the
difficulty in applying lattice methods to exclusive nonleptonic
decays.
\end{quote}
\vfill

\begin{center}
To appear in Heavy Flavours (2nd edition)\\
edited by A.J.~Buras and M.~Lindner\\
(World Scientific, Singapore)
\end{center}
\vfill

\begin{flushleft}
October 1997
\end{flushleft}
\newpage
% end of shep titlepage

\title{HEAVY QUARK PHYSICS FROM LATTICE QCD}
%\vspace{-3in}
\author{J.M.~Flynn and C.T.~Sachrajda}

\address{Department of Physics and Astronomy,
University of Southampton\\
Southampton SO17 1BJ, UK}

\maketitle

\abstracts{We review the application of lattice QCD to the
phenomenology of $b$- and $c$-quarks. After a short discussion of the
lattice techniques used to evaluate hadronic matrix elements and the
corresponding systematic uncertainties, we summarise results for
leptonic decay constants, $B$--$\bar B$ mixing, semileptonic and rare
radiative decays. A discussion of the determination of heavy quark
effective theory parameters is followed by an explanation of the
difficulty in applying lattice methods to exclusive nonleptonic
decays.}

\section{Introduction}

The lattice formulation of Quantum Chromodynamics (QCD) and large
scale numerical simulations on parallel supercomputers are enabling
theorists to evaluate the long-distance strong interaction effects in
weak decays of hadrons in general, and of those containing a heavy
($c$- or $b$-) quark in particular. The difficulty in quantifying
these non-perturbative QCD effects is the principal systematic error
in the determination of the elements of the Cabibbo-Kobayashi-Maskawa
(CKM) matrix from experimental measurements, and in using the data for
tests of the standard model of particle physics and in searches for
``new physics''. In this article we will review the main results
obtained from lattice simulations in recent years, and discuss the
principal sources of uncertainty and prospects for future
improvements. More details can be found, for example, in the review
talks presented at the annual symposia on lattice field theory (see
Refs.~\cite{lat921}$^{-}$\cite{onogi:lat97}) and in references
therein.

In the lattice formulation of quantum field theory space-time is
approximated by a discrete ``lattice'' of points and the physical
quantities of interest are evaluated numerically by computing the
corresponding functional integrals. For these computations to make
sense it is necessary for the lattice to be sufficiently large to
accommodate the particle(s) being studied ($L\gg 1$\,fm say, where $L$
is the spatial length of the lattice), and for the spacing between
neighbouring points ($a$) to be sufficiently small so that
perturbation theory can be used to interpolate between the lattice
points ($a\lqcd\ll 1$). The number of lattice points in a simulation
is limited by the available computing resources; current simulations
are performed with about 16--20 points in each spatial direction (up
to about 64 points if the effects of quark loops are neglected, i.e.\
in the so called ``quenched'' approximation). Thus it is possible to
work on lattices which have a spatial extent of about 2 fm and a
lattice spacing of 0.1 fm, perhaps satisying the above requirements.

Since the Compton wavelength of heavy quarks is small, one has to be
careful in their simulation. With a discretization of the Dirac
action, $\bar Q(x)(\Dslash + m_Q)Q(x)$, where $Q$ represents the field
of the quark and $m_Q$ its mass, the condition $a\lqcd\ll 1$ is not
sufficient, one also requires $am_Q\ll 1$. Currently available
lattices, in simulations performed in the quenched approximation, have
spacings in the range $1.5\gev < a^{-1} < 4\gev$, and so it is not
possible to simulate $b$-quarks directly in this way. Hence, one
approach to $B$-physics is to simulate hadrons with heavy quarks which
are still considered to be acceptably light (typically with masses in
the range of that of the charmed quark), and then to extrapolate the
results to the $b$-region, using scaling laws from the Heavy Quark
Effective Theory (HQET) where applicable.

An important alternative approach is to use the HQET, and to compute
physical quantities in terms of an expansion in inverse powers of the
mass of the $b$-quark (and $c$-quark where appropriate). The first
term in this expansion corresponds to a calculation using static
(infinitely) heavy quarks. The evaluation of the coefficients in this
expansion does not require propagating heavy quarks, and hence there
are no errors of $O(m_Qa)$. However, as will be discussed in more
detail in \secref{hqet}, the evaluation of the higher order terms in
this expansion requires the subtraction of power divergences (i.e.\
terms which diverge as inverse powers of the lattice spacing $a$),
which significantly reduces the precision which can be obtained.

In this article we focus on the decays of $b$- and  $c$-quarks. Another
important area of investigation is the spectroscopy of heavy quarkonia
(for a review of recent results and references to the original
literature see Ref.~\cite{cdavies}), in which the non-relativistic
formulation of lattice QCD~\cite{nrqcd} is generally used.

The plan of this article is as follows. In the remainder of this
Section we briefly review the general procedure used for evaluating
hadronic matrix elements in lattice simulations and discuss the
principal sources of uncertainty.  The following Sections contain the
status of the results for the leptonic decay constants $f_B$ and $f_D$
(\secref{fb}), the $B_B$-parameter of $B^0$-$\bar B^0$ mixing
(\secref{bb}), semileptonic decay amplitudes of $B$- and
$D$-mesons (\secref{sl}) and the parameters $\bar\Lambda$ (the
binding energy of the heavy quark in a hadron), $\lambda_1$ (its
kinetic energy) and $\lambda_2$ (the matrix element of the
chromomagnetic operator) of the HQET (Section~\ref{sec:hqet}).
Section~\ref{sec:hqet} also contains a general discussion of the
theoretical difficulties present in calculations of power corrections,
i.e.\ corrections of $O(1/m_Q)$, $O(1/m_Q^2)$ etc., in physical
quantities. Section~\ref{sec:exclusive} contains a brief summary of
the status of lattice calculations of exclusive non-leptonic decay
amplitudes, which is an area of investigation requiring considerable
theoretical development. Finally in Section~\ref{sec:concs} we present
a brief summary and outlook. \tabref{best} shows where our preferred
estimates for some of the more important physical quantities can be
found.

\begin{table}
\vspace{-8pt}
\caption[]{Where in this review to find results for some important
physical quantities.}
\label{tab:best}
\kern1em
\begin{center}
\begin{tabular}{>{\PBS\raggedright\hspace{0pt}}p{0.59\hsize}%
>{\PBS\raggedright\hspace{0pt}}p{0.36\hsize}}
\hline
\ts Quantity & See\dots \\[0.4ex]
\hline
\ts Leptonic decay constants: $\fd$, $\fds$, $\fb$, $\fbs$ \bs &
    Eqs.~(\ref{eq:fdbest}--\ref{eq:fbs/fbbest}) \bs \\
Renormalization-group-invariant $B$-parameter for $B$--$\bar B$
mixing, $\rgbb$ \bs & \eqref{bbbest} \bs \\
$\xi \equiv \fbs \rgbbs^{1/2}/\fbd \rgbbd^{1/2}$ \bs &
    \eqref{xibest} \bs \\
Form factors $f^+(0)$, $V(0)$, $A_1(0)$ and $A_2(0)$ for semileptonic
$D\to K,K^*,\pi,\rho$ decays \bs &
    \tabref{slDdecayresults} \bs \\
Form factor $f^+(0)$ and decay rate for $\btopi$ \bs &
     \tabref{btopi-results} (UKQCD values) \bs \\
Form factors $V(0)$, $A_1(0)$ and $A_2(0)$ and rate for $\btorho$ \bs &
     \tabref{btorho-results} (UKQCD values) \bs \\
Form factor $T(0) \equiv T_1(0) = iT_2(0)$ for $\btokstargamma$ &
     \tabref{btokstargamma} (UKQCD values) and \eqref{t0best} \\[0.4ex]
\hline
\end{tabular}
\end{center}
\end{table}

\subsection{Evaluation of Hadronic Matrix Elements}

By using the Operator Product Expansion, it is generally possible to
separate the short- and long-distance contributions to weak decay
amplitudes into Wilson coefficient functions and operator matrix
elements respectively. Thus in order to evaluate the non-perturbative
QCD effects it is necessary to compute the matrix elements of local
composite operators. This is achieved in lattice simulations, by the
direct computation of correlation functions of multi-local operators
composed of quark and gluon fields (in Euclidean space):
\begin{equation}
\langle0|\,O(x_1, x_2,\ldots,x_n)|0\rangle =\, \frac{1}{Z}
\int[DA_\mu][D\psi][D\bar\psi]\,e^{-S}\,O(x_1, x_2,\ldots,x_n)\ ,
\label{eq:vev}
\end{equation}
where $Z$ is the partition function 
\begin{equation}
Z=
\int[DA_\mu][D\psi][D\bar\psi]\,e^{-S}\ ,
\label{eq:zdef}
\end{equation}
$S$ is the action and the integrals are over quark and gluon fields
at each space-time point. In \eqref{vev} $O(x_1, x_2,\ldots,x_n)$
is a multi-local operator; the choice of $O$ governs the physics which 
can be studied. 

We now consider the two most frequently encountered cases, for which
$n{=}2$ or $3$. Let $O(x_1, x_2)$ be the bilocal operator 
\begin{equation}
O_2(x_1,x_2)=T\{J_h(x_1)J^\dagger_h(x_2)\}\ ,
\label{eq:o2}\end{equation}
where $J_h$ is an interpolating operator for the hadron $h$ whose
properties we wish to study. In lattice computations we evaluate the
two point correlation function
\begin{equation}
C_2(t_x)\equiv\sum_{\vec x}
\langle\,0\,|O_2(x, 0)|\,0\,\rangle\ .
\label{eq:c2def}\end{equation}
For sufficiently
large positive $t_x$ one obtains:
\begin{equation}
C_2(t_x)\simeq\frac{e^{-m_ht_x}}{2m_h}\,|\langle\,0\,|J_h(0)|\,h\,\rangle
|^2\ .
\label{eq:c2asymp}\end{equation}
In \eqref{c2def} $m_h$ is the mass of the hadron $h$, which is assumed
to be the lightest one which can be created by the operator
$J_h^\dagger$.  The contribution from each heavier hadron, $h'$ with
mass $m_{h'}$ say, is suppressed by an exponential factor,
$\exp\big(-(m_{h'}-m_h)t_x\big)$. In lattice simulations the
correlation function $C_2$ is computed numerically, and by fitting the
results to the expression in \eqref{c2asymp} both the mass $m_h$ and
the matrix element $\langle\,0\,|J_h(0)|\,h\,\rangle$ can be
determined. In this case the hadron $h$ is at rest, but of course it
is also possible to give it a non-zero momentum, $\vec p$ say, by
taking the Fourier transform in \eqref{c2def} with the appropriate
weighting factor $\exp(i\vec p\dotprod\vec x)$.

As an example consider the case in which $h$ is the $B$-meson and
$J_h$ is the axial current $A_\mu$ (with the flavour quantum numbers
of the $B$-meson). In this case one obtains the value of the leptonic
decay constant $f_B$,
\begin{equation}
\langle\,0\,|A_\mu(0)|\,B(p)\,\rangle = f_B\,p_\mu\ .
\label{eq:fbdef}\end{equation}

It will also be useful to consider three-point correlation functions:
\be
C_3(t_x, t_y) = \sum_{\vec x,\vec y}
e^{i\vec p\dotprod\vec x} e^{i\vec q\dotprod\vec y}
\< 0\,|\,J_2(\vec x, t_x)\, O(\vec y,t_y)\, J^\dagger_1(\vec 0, 0)\,
|\, 0\>\ ,
\label{eq:c3def}\ee
where, $J_1$ and $J_2$ are the interpolating operators for hadrons
$h_1$ and $h_2$ respectively, $O$ is a local operator, and we have assumed
that $t_x>t_y>0$. Inserting complete sets of states between
the operators in \eqref{c3def} we obtain
\beqn
\lefteqn{C_3(t_x, t_y) = 
\frac{e^{-E_1t_y}}{2 E_1}\ \frac{e^{-E_2(t_x - t_y)}}{2 E_2}\,
\< 0|J_2(\vec 0, 0)|h_2(\vec p, E_2)\>\times}\nn\\ 
& & \< h_2(\vec p, E_2)|O(\vec 0,0)|h_1(\vec p {+} \vec q, E_1)\>\,
\< h_1(\vec p {+} \vec q, E_1)|J^\dagger_1(\vec 0,0)|0\>+\cdots,
\label{eq:c3states}
\eeqn where $E_1=\sqrt{m_1^2 + (\vec p{+}\vec q)^2}$,
$E_2=\sqrt{m_2^2+\vec p^2}$ and the ellipsis represents the
contributions from heavier states. The exponential factors,
$\exp(-E_1t_y)$ and $\exp\big(-E_2(t_x-t_y)\big)$, assure that for
large time separations, $t_y$ and $t_x-t_y$, the contributions from
the lightest states dominate.  All the elements on the right-hand side
of \eqref{c3states} can be determined from two-point correlation
functions, with the exception of the matrix element $\<h_2|O|h_1\>$.
Thus by computing two- and three-point correlation functions the
matrix element $\<h_2|O|h_1\>$ can be determined.

The computation of three-point correlation functions is useful in
studying semileptonic and radiative weak decays of hadrons, e.g. if
$h_1$ is a $B$-meson, $h_2$ a $D$ meson and $O$ the vector current
$\bar{b}\gamma^\mu c$, then from this correlation function we
obtain the form factors relevant for semileptonic $B\to D$ decays.

We end this brief summary of lattice computations of hadronic matrix
elements with a word about the determination of the lattice spacing
$a$.  It is conventional to introduce the parameter
$\beta=6/g_0^2(a)$, where $g_0(a)$ is the bare coupling constant in
the theory with the lattice regularization. It is $\beta$ $\big($or
equivalently $g_0(a)$$\big)$ which is the input parameter in the
simulation, and the corresponding lattice spacing is then determined
by requiring that some physical quantity (which is computed in lattice
units) is equal to the experimental value. For example, one may
compute $m_\rho a$, where $m_\rho$ is the mass of the $\rho$-meson,
and determine the lattice spacing $a$ by dividing the result by
$769\mev$.

\subsection{Sources of Uncertainty in Lattice Computations}
\label{subsec:uncert}

Although lattice computations provide the opportunity, in principle, to
evaluate the non-perturbative QCD effects in weak decays of heavy
quarks from first principles and with no model assumptions or free
parameters, in practice the precision of the results is limited by the
available computing resources. In this section we outline the main
sources of uncertainty in these computations:
\begin{itemize}
\item{\em Statistical Errors:} The functional integrals in
Eq.~(\ref{eq:vev}) are evaluated by Monte-Carlo techniques. This leads
to sampling errors, which decrease as the number of field
configurations included in the estimate of the integrals is increased.
\item{\em Discretization Errors:} These are artefacts due to the
finiteness of the lattice spacing. Much effort is being devoted to
reducing these errors either by performing simulations at several
values of the lattice spacing and extrapolating the results to the
continuum limit ($a=0$), or by ``improving'' the lattice formulation
of QCD so that the error in a simulation at a fixed value of $a$ is
formally smaller~\cite{improvement}$^-$\cite{hsmpr}. In particular, it has
recently been shown to be possible to formulate lattice QCD in such a
way that the discretization errors vanish quadratically with the
lattice spacing~\cite{luscher}, even for non-zero quark
masses~\cite{mrsstt}.  This is in contrast with the traditional Wilson
formulation~\cite{wilson}, in which the errors vanish only linearly.
Some of the simulations discussed in the following sections have used
a tree-level improved action and operators; in these studies the
artefacts formally vanish more quickly $\big($like $a
\alpha_s(a)$$\big)$ than for the Wilson action. In the following, this
tree-level improved action will be denoted as the SW (after
Sheikholeslami-Wohlert who first proposed it~\cite{sw}) or ``clover''
action.
\item{\em Extrapolations to Physical Quark Masses:} It is generally
not possible to use physical quark masses in simulations. For the
light ($u$- and $d$-) quarks the masses must be chosen such that the
corresponding $\pi$-mesons are sufficiently heavy to be insensitive to
the finite volume of the lattice. In addition, as the masses of the
quarks are reduced the computer time necessary to maintain the
required level of precision increases rapidly. For the heavy quarks
$Q$ (i.e.\ for $c$, and particularly for $b$) the masses must be
sufficiently small so that the discretization errors, which are of
$O(m_Q a)$ or $O(m_Q^2a^2)$, are small. The results obtained in the
simulations must then be extrapolated to those corresponding to
physical quark masses.
\item{\em Finite Volume Effects:} We require that the results we
obtain be independent of the size of the lattice. Thus, in principle,
the spatial size of the lattice $L$ should be $\gg 1$\,fm (in practice
$L\gtrsim 2$--3\,fm), and as mentioned above, we cannot use very light
$u$- and $d$-quarks (in order to avoid very light pions whose
correlation lengths, i.e.\ the inverses of their masses, would be of
$O(L)$ or greater).
\item{\em Contamination from Excited States:} These are the
uncertainties due to the effects of the excited states, represented by
the ellipsis in Eq.~(\ref{eq:c2asymp}). In most simulations, this is
not a serious source of error (it is, however, more significant in
computations with static quarks).  Indeed, by evaluating correlation
functions $\langle J_h(x)\,J'_h(0)\rangle$ for a variety of
interpolating operators $\{J_h, J'_h\}$, it is possible to obtain the
masses and matrix elements of the excited hadrons (for a recent
example see~\cite{excited}).
\item{\em Lattice-Continuum Matching:} The operators used in lattice
simulations are bare operators defined with the lattice spacing as the
ultra-violet cut-off. From the matrix elements of the bare lattice
composite operators, we need to obtain those of the corresponding
renormalized operators defined in some standard continuum
renormalization scheme, such as the $\overline\mathrm{MS}$ scheme. The
relation between the matrix elements of lattice and continuum
composite operators involves only short-distance physics, and hence
can be obtained in perturbation theory. Experience has taught us,
however, that the coefficients in lattice perturbation theory can be
large, leading to significant uncertainties (frequently of $O(10\%)$
or more).  For this reason, non-perturbative techniques to evaluate
the renormalization constants which relate the bare lattice operators
to the renormalized ones have been developed, using chiral Ward
identities where possible~\cite{ward} or by imposing an explicit
renormalization condition~\cite{npren} (see also
Refs.~\cite{luscher,luscher2}), thus effectively removing this source
of uncertainty in many important cases.
\begin{figure}
\hbox to\hsize{\hfill
\epsfxsize0.68\hsize\epsffile{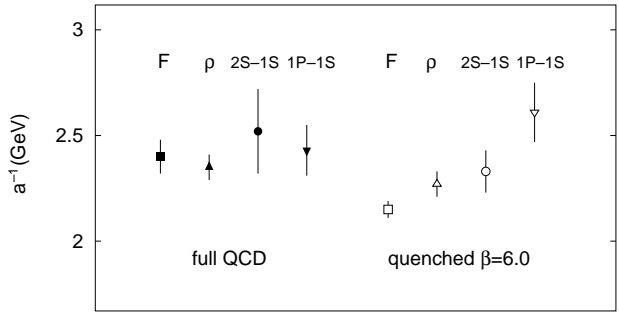}
\hfill}
\caption[]{Comparison of lattice spacings (the quantity plotted is
actually the \emph{inverse} lattice spacing, $a^{-1}$) determined from
different physical quantities in quenched and unquenched simulations
by the SESAM and T$\chi$L collaborations~\cite{henning}. $F$ denotes
the value determined from the static quark potential and $\rho$
denotes the value determined from the $\rho$-meson mass, while
$2S$--$1S$ and $1P$--$1S$ denote values obtained from energy level
splittings in $Q \bar Q$ bound states using the lattice formulation of
nonrelativistic QCD.}
\label{fig:sesamtxl}
\end{figure}
\item{\em ``Quenching'':} In most current simulations the matrix
elements are evaluated in the ``quenched'' approximation, in which the
effects of virtual quark loops are neglected. For each gluon
configuration $\{U_\mu(x)\}$, the functional integral over the quark
fields in Eq.~(\ref{eq:vev}) can be performed formally, giving the
determinant of the Dirac operator in the gluon background field
corresponding to this configuration. The numerical evaluation of this
determinant is possible, but is computationally very expensive, and
for this reason the determinant is frequently set equal to its average
value, which is equivalent to neglecting virtual quark
loops. Gradually, however, unquenched calculations are beginning to be
performed, e.g.\ in \figref{sesamtxl} we show the lattice
spacing obtained by the SESAM and T$\chi$L collaborations from four
physical quantities in both quenched and unquenched
simulations~\cite{henning}. In the quenched case there is a spread of
results of about $\pm 10\%$, whereas in the unquenched case the spread
is smaller (although the errors are still sizeable for some of the
quantities used).  In the next 3--5 years it should be possible to
compute most of the physical quantities discussed below without
imposing this approximation.
\end{itemize}

\section{The Leptonic Decay Constants ${f_D}$ and ${f_B}$}
\label{sec:fb}

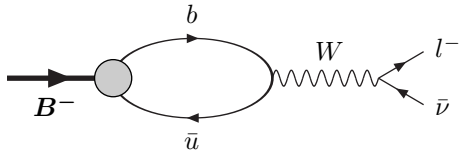
\begin{figure}
\begin{center}
\begin{picture}(180,50)(0,15)
\SetWidth{2}\ArrowLine(10,41)(50,41)
\SetWidth{0.5}
%\Line(10,40)(50,40)\Line(10,42)(50,42)
\Oval(80,41)(15,30)(0)
\ArrowLine(79,56)(81,56)\ArrowLine(81,26)(79,26)
\GCirc(50,41){7}{0.8}
\Photon(110,41)(150,41){3}{7}
\ArrowLine(150,41)(167,51)\ArrowLine(167,31)(150,41)
\Text(20,35)[tl]{\boldmath$B^-$}
\Text(80,62)[b]{$b$}\Text(80,20)[t]{$\bar u$}
\Text(172,53)[l]{$l^-$}\Text(172,30)[l]{$\bar\nu$}
\Text(132,48)[b]{$W$}
\end{picture}
\caption{Diagram representing the leptonic decay of the $B$-meson.}
\label{fig:fb}
\end{center}
\end{figure}

In this section we review the current status of calculations of the
leptonic decay constants $f_D$ and $f_B$.  Leptonic decays of heavy
mesons, see \figref{fb}, are particularly simple to treat
theoretically~\footnote{For simplicity, the presentation here is for
pseudoscalar mesons $D$ and $B$. A parallel discussion holds also for
the vector mesons $D^*$ and $B^*$.}. In each case the strong
interaction effects are contained in a single parameter,
called the decay constant, $f_D$ or $f_B$.  Parity symmetry implies
that only the axial component of the $V$--$A$ weak current contributes
to the decay, and Lorentz invariance that the matrix element of the
axial current is proportional to the momentum of the meson (with the
constant of proportionality defined to be the decay constant) as, for
example, in \eqref{fbdef}.

Knowledge of $f_B$ would allow us to predict the rates for the
corresponding decays:
\begin{equation}
\Gamma(B\to l\nu_l + l\nu_l \gamma) = 
\frac{G_F^2 V_{ub}^2}{8\pi}f_B^2m_l^2m_B
\Big(\!1 - \frac{m_l^2}{m_B^2}\Big)^{\!2} \big( 1 + O(\alpha)\big)\; ,
\label{eq:rate}\end{equation}
where the $O(\alpha)$ corrections are also known. The value of $f_B$
is very important in describing $B$--$\bar B$ mixing as explained
in~\secref{bb}.  Knowing $f_B$ would also be useful for our
understanding of other processes in $B$-physics, particularly those
for which ``factorization'' turns out to be a useful approximation.

\begin{table}
\vspace{-8pt}
\caption[]{Results for $\fd$ and $\fds$ using the conventional
  formulation of heavy quarks. All values were obtained in the
  quenched approximation, except for those from the HEMCGC
  collaboration which were computed for two flavours of sea quarks
  ($n_f=2$) (we have combined quoted systematic errors from HEMCGC in
  quadrature). Infinite values of $\beta$ symbolise results obtained
  after extrapolation to the continuum limit. The error quoted by APE
  includes an estimate of discretization uncertainties taken from a
  comparision of results at $\beta=6.0$ and $6.2$. Results from MILC
  and JLQCD are preliminary.}
\label{tab:fdfds}
\kern1.5em
\hbox to\hsize{\hss
\begin{tabular}{@{}rcccllll@{}}
\hline
\ts & Yr & $\beta$ & $\csw$ & norm &
 $\fd/\!\mev$ & $\fds/\!\mev$ & $\fds/\fd$ \\[0.4ex]
\hline
\ts \small{MILC}~\cite{milc:fb-lat97} & 97 & $\infty$ & 0 & nr &
  186(10)\errp{27}{18}\errp90 & 199(8)\errp{40}{11}\errp{10}0 &
 1.09(2)\errp51\errp20
\\
\small{JLQCD}~\cite{jlqcd:fb-lat97} & 97 & $\infty$ & 0\rlap{,1} & nr &
  192(10)\errp{11}{16} & 213(11)\errp{12}{18} \\
\small{APE}~\cite{ape:fb97} & 97 & 6.2 & 1 & rel &
 221(17) & 237(16) & 1.07(4) \\
\small{LANL}~\cite{lanl:fb96} & 96 & 6.0 & 0 & nr &
  229(7)\errp{20}{16} & 260(4)\errp{27}{22} & 1.135(2)\errp6{23} \\
\small{LANL}~\cite{lanl:fb96} & 96 & $\infty$ & 0 & nr &
  186(29) & 218(15) \\
\small{PCW}~\cite{pcw:fb94} & 94 & $\infty$ & 0 & nr &
  170(30) & & 1.09(2)(5) \\
\small{UKQCD}~\cite{ukqcd:fb94} & 94 & 6.2 & 1 & rel &
  185\errp{4}{3}\errp{42}{7} & 212\errp{4}{4}\errp{46}{7} & 1.18(2) \\
\small{UKQCD}~\cite{ukqcd:fb94} & 94 & 6.0 & 1 & rel &
 199\errp{14}{15}\errp{27}{19} 
  & 225\errp{15}{15}\errp{30}{22} & 1.13\errp67 \\
\small{BLS}~\cite{bls:fb} & 94 & 6.3 & 0 & nr &
  208(9)(35)(12) & 230(7)(30)(18) & 1.11(6) \\
\small{HEMCGC}~\cite{hemcgc:fb94}& 94 & 5.3 & 0 & nr &
  215(5)(53) & 287(5)(60) \\
%HEMCGC}~\cite{hemcgc:fb94}& 94 & 5.3 & 0 & nr &
%  215(5)(40)(35)(5) & 287(5)(45)(40)(5) \\
\small{HEMCGC}~\cite{hemcgc:fb93} & 93 & 5.6 & 0 & &
  200--287 & 220--320 \\
\small{ELC}~\cite{elc:fb92} & 92 & 6.4 & 0 & rel & 210(40) & 230(50) \\
\small{PCW}~\cite{pcw:fb91} & 91 & 6.0 & 0 & rel &
  198(17) & 209(18) \\
\small{ELC}~\cite{elc:fd88} & 88 & 6.2 & 0 & rel & 181(27) & 218(27) \\
\small{ELC}~\cite{elc:fd88} & 88 & 6.0 & 0 & rel & 197(14) & 214(19) \\
\small{DeGL}~\cite{degrandloft} & 88 & 6.0 & 0 & rel &
  190(33) & 222(16) & 1.17(22) \\
\small{BDHS}~\cite{bdhs} & 88 & 6.1 & 0 & rel &
 174(26)(46) & 234(46)(55) & 1.35(22) \\[0.4ex]
\hline
\end{tabular}
\hss}
\end{table}

Lattice results for decay constants of charmed and bottom mesons
obtained over the last ten years are summarised in
Tables~\ref{tab:fdfds} and \ref{tab:fbfbs}. These are presented using
a normalization in which $f_{\pi^+}\simeq 131\mev$.

Since we are dealing with a heavy quark with mass $m_Q$, the product
$m_Qa$ is large and can be an important source of mass-dependent
discretization errors. This has been addressed in two ways in the
decay constant calculations from 1994 onwards. Some studies, denoted
by $\csw=1$, have used the Sheikholeslami-Wohlert (SW or clover)
improved action~\cite{sw} together with improved
operators~\cite{hsmpr}, which removes tree level $O(a)$ errors,
leaving ones of $O(\alpha_s m_Qa)$. In the near future we can look
forward to the application of recently developed techniques to reduce
the discretization errors still further, to ones of $O(m_Q^2
a^2)$~\cite{luscher,mrsstt}. The second approach introduces a revised
normalization of the quark fields in simulations using the standard
Wilson fermion action.  This is designed to remove higher order
effects in $m_Qa$ from the heavy quark propagator, but only at tree
level in the strong interactions, and is distinguished in the table by
the label `nr', denoting a nonrelativistic normalization (in contrast
to the standard `rel' or relativisitic one). The theoretical
significance of this normalization factor is not completely clear; in
particular, it does not eliminate all the $O(m_Qa)$ effects.  The
nonrelativistic normalization is often denoted `KLM' in the
literature, after some of its
originators~\cite{klm:mack,klm:kron}$^-$\cite{klm:ekm}.  For more
details of the lattice formulations and improvement procedures, see
the recent review by Wittig~\cite{how:ijmp-review}.

\begin{table}
\vspace{-8pt}
\caption[]{Results for $\fb$ and $\fbs$ using the conventional
  formulation of heavy quarks. All values were obtained in the
  quenched approximation, except for those from the HEMCGC
  collaboration which were computed with two flavours of sea quarks
  ($n_f=2$) (we have combined quoted systematic errors from HEMCGC in
  quadrature). Infinite values of $\beta$ symbolise results obtained
  after extrapolation to the continuum limit. The error quoted by APE
  includes an estimate of discretization uncertainties taken from a
  comparision of results at $\beta=6.0$ and $6.2$. Results from MILC
  and JLQCD are preliminary.}
\label{tab:fbfbs}
\kern1.5em
\hbox to\hsize{\hss
\begin{tabular}{@{}rcccllll@{}}
\hline
\ts  & Yr & $\beta$ & $\csw$ & norm &
 $\fb/\!\mev$ & $\fbs/\!\mev$ & $\fbs/\fb$ \\[0.4ex]
\hline
\ts \small{MILC}~\cite{milc:fb-lat97} & 97 & $\infty$ & 0 & nr &
  153(10)\errp{36}{13}\errp{13}0 & 164(9)\errp{47}{13}\errp{16}0 &
  1.10(2)\errp53\errp32
\\
\small{JLQCD}~\cite{jlqcd:fb-lat97} & 97 & $\infty$ & 0\rlap{,1} & nr &
  163(12)\errp{13}{16} & 180(16)\errp{14}{18} \\
\small{APE}~\cite{ape:fb97} & 97 & 6.2 & 1 & rel & 180(32) & 205(35) & 1.14(8)
\\
\small{PCW}~\cite{pcw:fb94} & 94 & $\infty$ & 0 & nr & 
  180(50) &  & 1.09(2)(5) \\
\small{UKQCD}~\cite{ukqcd:fb94} & 94 & 6.2 & 1 & rel &
  160(6)\errp{59}{19} & 194\errp65\errp{62}9 & 1.22\errp43 \\
\small{UKQCD}~\cite{ukqcd:fb94} & 94 & 6.0 & 1 & rel &
 176\errp{25}{24}\errp{33}{15} & & 1.17(12) \\
\small{BLS}~\cite{bls:fb} & 94 & 6.3 & 0 & nr &
  187(10)(34)(15) & 207(9)(34)(22) & 1.11(6) \\
%HEMCGC}~\cite{hemcgc:fb94} & 94 & 5.3 & 0 & nr &
%  150(10)(40)(40)(5) \\
\small{HEMCGC}~\cite{hemcgc:fb94} & 94 & 5.3 & 0 & nr &
  150(10)(57) \\
\small{HEMCGC}~\cite{hemcgc:fb93} & 93 & 5.6 & 0 & & 152--235 \\
\small{ELC}~\cite{elc:fb92} & 92 & 6.4 & 0 & rel & 205(40) \\
\small{BDHS}~\cite{bdhs} & 88 & 6.1 & 0 & rel & 105(17)(30) \\[0.4ex]
\hline
\end{tabular}
\hss}
\end{table}

\begin{figure}
\hbox to\hsize{\hss
\epsfxsize0.7\hsize\epsffile{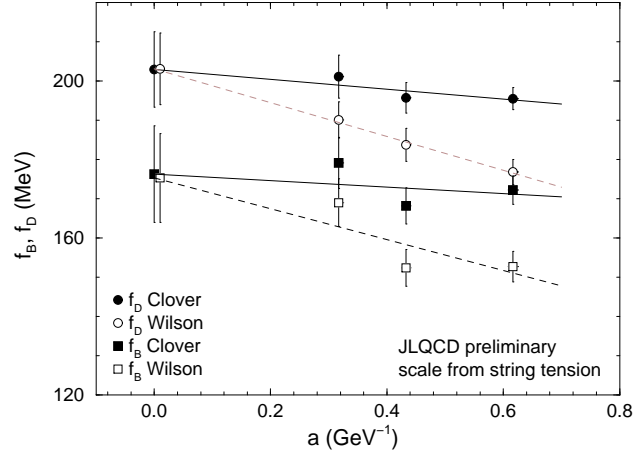}
\hss}
\caption[]{Continuum extrapolation of JLQCD~\cite{jlqcd:fb-lat97}
results (preliminary) for $f_D$ and $f_B$. The graph shows agreement
between results from two different lattice formulations when
extrapolated to zero lattice spacing. Open symbols denote an
unimproved Wilson action ($\csw=0$), filled symbols denote an improved
Sheikholeslami-Wohlert or Clover ($\csw=1$) action.}
\label{fig:jlqcd-fdfb}
\end{figure}

The main efforts of the lattice community are being devoted to
controlling the systematic uncertainties, which dominate the
errors. This is illustrated by the latest, and still preliminary,
results, which come from the MILC~\cite{milc:fb-lat97} and
JLQCD~\cite{jlqcd:fb-lat97} collaborations.  These collaborations are
carrying out extensive simulations using many different lattice
spacings to allow them to extrapolate to the continuum limit. JLQCD
study different prescriptions for reducing the $O(m_Qa)$
discretization errors associated with heavy quarks with the aim of
demonstrating that all results converge in the continuum limit. Their
error is statistical combined with the uncertainty due to the spread
over prescriptions.  In \figref{jlqcd-fdfb} we show the continuum
extrapolation of their results for $f_D$ and $f_B$ for two different
lattice fermion formulations. In this case the continuum values of
$f_B$ and $f_D$ obtained using the two formulations agree remarkably
well; the agreement is still acceptable (although not so remarkable)
when quantities other than the string tension are used to determine
the lattice spacing. We also note that in this case the dependence on
the lattice spacing is milder for the improved action as expected 
(although further studies are needed to confirm whether this is a 
general feature).

\begin{figure}
\epsfxsize0.7\hsize \unit\epsfxsize
\hbox to\hsize{\hss\vbox{\offinterlineskip
\epsffile[-23 54 291 231]{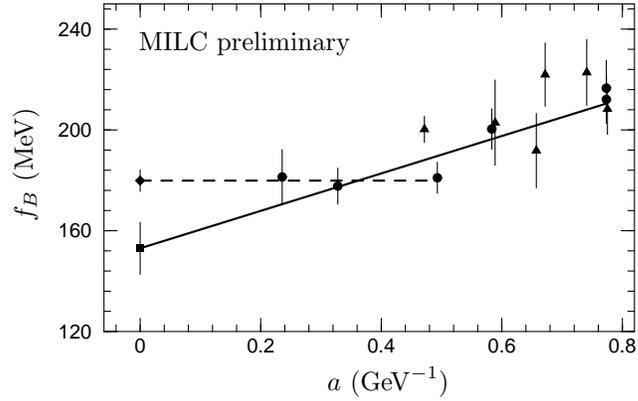}
\point 0.2 0.52 {\hbox{MILC preliminary}}
\point 0 0.4 {\begin{sideways}$f_B\ (\!\mev$)\end{sideways}}
}\hss}
\hbox to\hsize{\hss
\hbox to\unit{\kern0.5\unit$a\ (\!\gev^{-1})$\hfill}\hss}
\caption[]{Continuum extrapolation of MILC~\cite{milc:fb-lat97}
results (preliminary) for $f_B$. Circles (triangles) denote results
from quenched (unquenched $n_f=2$) simulations. The solid line and
square denote the continuum extrapolation by a linear fit in the
lattice spacing $a$ to the quenched results, while the dashed line and
diamond denote a weighted average of the three quenched results at
smallest $a$. The scale is set by $f_\pi$.}
\label{fig:milc-fb}
\end{figure}

MILC simulate a range of heavy quark masses, giving meson masses
straddling $m_D$, together with a static (infinite mass) quark,
allowing an interpolation to $m_B$. The continuum extrapolation of
their results is illustrated in \figref{milc-fb}. In the MILC results,
the first error is statistical and the second is systematic within the
quenched approximation.  MILC also have some unquenched simulation
results (with $n_f=2$ dynamical flavours). Their final error is for
the effects of quenching, which they estimate from: (i) the difference
between the unquenched results at the smallest available lattice
spacing and the quenched results interpolated to the same spacing, and
(ii) comparing the results when the lattice spacing is fixed by
$f_\pi$ or $m_\rho$.\footnote{For $\fds$ and $\fbs$ they also compare
results where the strange quark mass is determined by $m_K$ or
$m_\phi$.} Their results suggest that unquenching may raise the value
of the decay constants. This agrees with
estimates~\cite{sharpe-zhang,sharpe:lat96}, using the difference
between chiral loop contributions in quenched and unquenched QCD, that
$\fb$ and $f_{B_s}$ in full QCD are increased by 10--20\% over their
quenched values. Lattice simulations are possible using scalar fields
with a quark-like action which act effectively as negative numbers of
ordinary quark flavours~\cite{bermions3}. Calculations of $f_B$ in the
static limit extrapolated from negative numbers of flavours also
suggest an increase of about 20\%.

At this point we feel we need to add a word of caution. In their
evaluation of $f_B$, both the JLQCD and MILC collaborations simulate
heavy quarks with masses $m_Q$ satisfying $m_Qa > 1$ and apply a
nonrelativistic or `KLM' normalization prescription. Even with such a
normalization, however, one should worry about the discretization
errors when such large masses are used. Not only would one expect
these errors to be sizeable at a fixed value of beta, but the
extrapolation to the continuum limit is now complicated since all
powers of $m_Q a$ contribute significantly to the results which are
being extrapolated. Set against these worries is the observation by
the MILC collaboration that they obtain very similar values for $f_B$
if they restrict the set of points to be extrapolated to those with
heavy quarks with reasonably small masses (the differences are
included in their error bars).  Our own strong preference is to
restrict the simulations to heavy quarks with masses significantly
smaller than~1 in lattice units (and to use improved actions).

In obtaining a continuum value for $f_B$ one has to decide in what
order to perform the extrapolations in $a$ and $m_Q$.  The MILC
collaboration obtain $f_B$ by fitting, at each finite lattice spacing
$a$, the value of the decay constant of a static quark (infinite mass)
and those of several mesons with finite masses (it should be noted
that these have different systematic errors). They then extrapolate
the results for $f_B$ to the continuum limit to obtain their final
result.  An alternative
procedure~\cite{pcw:fb94,how:ijmp-review,sommer} is to perform
separately the continuum extrapolations of the static results and
results obtained at fixed finite meson masses. The mass dependence of
the continuum points is then fitted to determine the physical decay
constants. The most recent analysis using this method has been
performed by Wittig~\cite{how:ijmp-review}, combining lattice data
from many simulations. The MILC and JLQCD data is excluded, since it
is still preliminary, and the results obtained with the clover action
have also not been included since it was felt that there is
insufficient data to perform the continuum extrapolation.  This
analysis shows some evidence for discretization errors at large quark
masses, where results using different normalization procedures, mass
definitions and so on, begin to diverge. Wittig's final fit,
therefore, is to the static point and to points with meson masses less
than about $2\gev$. The results of this procedure
give~\cite{how:ijmp-review}
\begin{eqnarray}
\fd  &=&  191 \err{19}{28} \mev \\
\fds &=&  206 \err{18}{28} \mev \\
\fb  &=&  172 \err{27}{31} \mev
\end{eqnarray}
The statistical error and the systematic error arising from fixing the
lattice spacing are the dominant (and comparable) uncertainties in
these numbers. The central values are obtained using the continuum
extrapolated value of $f_\pi$ to fix the lattice
spacing~\cite{how:ijmp-review}.

The dimensionless ratios $\fds/\fd$ and $\fbs/\fb$ are also of great
interest and have the advantage that many systematic effects should
partially cancel. The lattice results illustrate, as expected, that
the value of the decay constant decreases as the mass of the light
valence quark decreases. Here a straightforward extrapolation of world
results by Wittig~\cite{how:ijmp-review} gives
\begin{eqnarray}
\fds/\fd &=& 1.08 \pm 0.08, \\
\fbs/\fb &=& 1.14 \pm 0.08,
\end{eqnarray}
which are compatible with the preliminary MILC
results~\cite{milc:fb-lat97}. These results are obtained in the
quenched approximation. MILC estimate the quenching error in the
ratios to be about 3\%.  One would expect a small correction here from
cancellation of the effects in $\fbs$ and $\fb$ separately. We note
that the chiral loop estimate for the difference of the ratio
$\fbs/\fb$ from 1 can be surprisingly
large~\cite{sharpe-zhang,sharpe:lat96}, e.g. for one choice of
parameters it can be as large as ${+}16\%$.

\subsection{Summary and Conclusions}

We now summarise this rather lengthy discussion and present our
conclusions.  The two principle sources of uncertainty in the
calculations of the decay constants are those due to discretization
errors and to quenching.  The effects of the latter are only now
beginning to be studied, and early indications suggest that the values
of the decay constants may increase when the effects of quark loops are
fully included, but this still needs to be confirmed. The
discretization errors are currently studied either by using improved
actions and operators, or by performing the computations at a sequence
of lattice spacings and extrapolating the results to the continuum
limit. The results from the two approaches are consistent, and further
confirmation that these errors are under control will come after the
implementation of the $O(a^2)$  improvement techniques  proposed in
Refs.~\cite{luscher,mrsstt}. Our view of the current status of the
calculations can be summarised by the following values for the decay
constants and their ratios: 
\begin{eqnarray}
\fd & = & 200 \pm 30\mev\label{eq:fdbest}\\ 
\fds & = & 220 \pm 30\mev\label{eq:fdsbest}\\ 
\fb & = & 170 \pm 35\mev\label{eq:fbbest}\\
\fbs & = & 195 \pm 35\mev\label{eq:fbsbest}\\ 
\fds/\fd & = & 1.10 \pm 0.06 \label{eq:fds/fdbest}\\  
\fbs/\fb & = & 1.14 \pm 0.08 \label{eq:fbs/fbbest}  
\end{eqnarray}
The principal difficulty in presenting results for physical quantities
such as those in Eqs.~(\ref{eq:fdbest}--\ref{eq:fbs/fbbest}) is to
estimate the errors. The value of the lattice spacing typically varies
by 10\% or so depending on which physical quantity is used to
calibrate the lattice simulation. This variation is largely due to the
use of the quenched approximation. We therefore consider that
$\pm10\%$ is an irreducible minimum error in decay constants computed
in quenched simulations (15\% when they are computed in the static
limit, since in that case it is $f_B\sqrt{m_B}$ which is computed
directly). The remainder of each error in
Eqs.~(\ref{eq:fdbest}--\ref{eq:fbs/fbbest}) and below is based on our
estimates of the other uncertainties, particularly those due to
discretization errors and the normalization of the lattice composite
operators. As explained in the Introduction, much successful work is
currently being done to reduce these uncertainties.

It is very interesting to compare the lattice \emph{prediction} for
$\fds$ in \eqref{fdsbest} with experimental measurements. Combining
four measurements of $\fds$ from $D_s\to\mu\nu$ decays, the rapporteur
at the 1996 ICHEP conference found~\cite{richman}
\begin{equation}
f_{D_s} = 241 \pm 21 \pm 30 \mev .
\label{eq:richman}
\end{equation}
In spite of the sizeable errors, the agreement with the lattice
prediction is very pleasing and gives us further confidence in the
predictions for $f_B$ and related quantities.

We end this section with a comment on the validity of the 
asymptotic scaling law for the decay constants. 
For sufficiently large masses of the heavy quark, the decay constant
of a heavy--light pseudoscalar meson $P$ scales with its mass $m_P$ as
follows:
\begin{equation}
f_P = \frac{A}{\sqrt{m_P}}\left[\alpha_s(m_P)^{-2/\beta_0}\left\{1 + 
O\big(\alpha_s(m_P)\big)\right\} + O\left(\frac{1}{m_P}\right)\right] ,
\label{eq:fpscaling}
\end{equation}
where $A$ is independent of $m_P$.  Using the leading term of this
scaling law, a value of $200\mev$ for $f_D$ would correspond to
$f_B\simeq 120\mev$. Results from the lattice computations shown
in~\tabref{fbfbs}, however, indicate that $f_B$ is significantly
larger than this and that the $O(1/m_P)$ corrections on the right-hand
side of \eqref{fpscaling} are considerable.  The coefficient of the
$O(1/m_P)$ corrections is typically found to be between $0.5$ and
$1\gev$, on the large side of theoretical expectations.

\section{$B^0-\bar B^0$ Mixing}
\label{sec:bb}

$B$--$\bar{B}$ mixing provides important constraints for the
determination of the angles of the unitarity triangle (for a review
and references to the original literature see e.g. Ref.~\cite{bfl}).
In this process, the strong interaction effects are contained in the
matrix element of the $\Delta B{=}2$ operator:
\begin{equation}
M(\mu) = \langle\bar B^0 |\, \bar b \gamma_\mu (1{-}\gamma_5) q\,
  \bar b \gamma^\mu (1{-}\gamma_5) q \,| B^0 \rangle, 
\label{eq:mbbar}
\end{equation}
where $q {=} d$ or $s$ (unless indicated otherwise, we assume that
$q{=}d$). The argument $\mu$ implies that the operator has been
renormalized at the scale $\mu$. It is conventional to introduce the
$\bb$-parameter through the definition
\begin{equation}
M(\mu) = \frac{8}{3}\, f_B^2 m_B^2 \bb(\mu).
\label{eq:bbdef}
\end{equation}
The dimensionless quantity $\bb$ is better-determined than $f_B$ in
lattice calculations, so that the theoretical uncertainties in the
value of the matrix element $M$, needed for phenomenology,
are dominated by our ignorance of $f_B$.

$\bb(\mu)$ is a scale dependent quantity for which lattice results are
most often quoted after translation to the $\overline\mathrm{MS}$
scheme. The next-to-leading order (NLO) renormalization group
invariant $B$--parameter ($\rgbb^{\mathrm{nlo}}$) is defined by
\begin{equation}
\rgbb^{\mathrm{nlo}} = \alpha_s(\mu)^{-2/\beta_0}
 \left( 1 + {\alpha_s(\mu)\over4\pi} J_{n_f} \right) \bb(\mu),
\label{eq:bbhat}
\end{equation}
where $\beta_0 = 11 - 2n_f/3$ and $J_{n_f}$ is obtained from the
one- and two-loop anomalous dimensions of the $\Delta B{=}2$ operator
by~\cite{bjw:bb},
\begin{equation}
J_{n_f} = {1\over2\beta_0} \left( \gamma_0 {\beta_1\over\beta_0} -
 \gamma_1 \right),
\end{equation}
with $\beta_1 = 102 - 38 n_f/3$, $\gamma_0 = 4$ and $\gamma_1 = -7 + 4
n_f/9$. In the discussion below we choose $\mu=m_b$.

Results for $\bb$ are summarised in \tabref{bbstat} for static $b$
quarks and in \tabref{bbprop} for propagating $b$-quarks. We now discuss
these two cases in turn. 

\subsection{$\bb$ obtained using static $b$-quarks}

\begin{table}
\vspace{-8pt}
\caption[]{Results for the mixing parameter $\bb$ obtained in the
quenched approximation using static $b$-quarks (the value for $\csw$
refers to the light quark action). $\bb(m_b)$ with $m_b=5\gev$ is
calculated from the raw lattice matrix elements and then converted to
$\rgbb^{\mathrm{nlo}}$, the renormalization group invariant $B$
parameter. For ease of comparison we have used the same choice of
parameters in all cases together with NLO matching from lattice to
continuum. Where this has changed the results from those quoted by the
authors, the result obtained by us is indicated in oblique type. M1,
M2 and M3 denote three different but equivalent ways of organising the
matching calculations (see text).}
\label{tab:bbstat}
\kern1em
\begin{center}
\begin{tabular}{rccccll}
\hline
\ts  & Yr & $\beta$ & $\csw$ & & $\bb(m_b)$ & $\rgbb^{\mathrm{nlo}}$
 \\[0.4ex]
\hline
\ts                 &    &     &   & M1 & 0.76(5) & 1.21(8) \\
GM~\cite{gm:static} & 97 & 6.0 & 1 & M2 & 0.54(4) & 0.86(6) \\
                    &    &     &   & M3 & \sl0.86(5) & \sl1.37(8) \\[0.8ex]
                    &    &     &   & M1 & \sl0.76(6) & \sl1.20(9) \\
UKQCD~\cite{ukqcd:staticb} & 96 & 6.2 & 1 & M2 & \sl0.57(6) & \sl0.91(9) \\
                    &    &     &   & M3 & \sl0.82(6) & \sl1.31(9) \\[0.8ex]
                    &    &     &   & M1 & \sl0.93(4) & \sl1.48(7) \\
Ken~\cite{ken:bb}   & 96 & 6.0 & 0 & M2 & \sl0.75(4) & \sl1.19(7) \\
                    &    &     &   & M3 & \sl0.98(4) & \sl1.56(7) \\[0.4ex]
\hline
\end{tabular}
\end{center}
\end{table}

The main difficulty in trying to determine the value of $\bb$ using
static heavy quarks is due to the large perturbative corrections which
are encountered when calculating the matrix element $M(\mu)$ in some
standard renormalization scheme (e.g. the $\MSbar$ scheme) in full QCD
from those measured on the lattice in the effective theory. We will
illustrate this below. Because of the relatively large uncertainties 
which result from the truncation of the perturbation series, we
believe that, at present, the calculations of $\bb$ in the static theory
add little to the information obtained with propagating quarks.  For
this reason our ``best'' lattice value for $\bb$ is obtained using
results obtained with propagating quarks only, and will be presented in
section~\ref{subsec:bbprop} below.

In order to compute $M(m_b)$ one needs to evaluate the matrix elements
of four $\Delta B=2$ operators in the static lattice theory 
(denoted by $O_i(a)$ with $i=L,N,R,S$). The relation between $M(m_b)$
and the matrix elements of the $O_i$ is of the form:
\begin{equation}
M(m_b)=\sum_{i=L,N,R,S} Z_i(m_b, a)\, \langle O_i(a) \rangle\ .
\label{eq:bbmatching}\end{equation}
The explicit form of the operators and details of the matching can be found
in Refs.~\cite{fhh}$^-$\cite{buchalla:nlo}. The matching can be 
performed in three stages:
\begin{itemize}
\item[(i)] From the matrix elements of the lattice operators in the
effective theory one can obtain the corresponding ones renormalized in
some standard continuum scheme at a scale $\mu=a^{-1}$.
\item[(ii)] Using the renormalization group equations one then obtains
the continuum effective theory matrix elements of operators renormalized
at a scale $\mu=m_b$.
\item[(iii)] Finally, by matching the continuum effective theory and
full QCD one obtains $M(m_b)$.
\end{itemize}
We illustrate this procedure by considering $Z_L$, which in practice
is the most complicated case (the operator $O_L$ is just that in 
Eq.~(\ref{eq:mbbar}) but with static heavy quarks). 
It can be written in the form:
\begin{eqnarray}
Z_L & = & \bigg\{C_L(m_b)\aalpha^{d_1}\left(1 + 
\frac{\alpha_s(a^{-1}) -\alpha_s(m_b)}{4\pi}J\right) +\nonumber\\ 
& &\hspace{-30pt}
C_S(m_b)\bigg[\aalpha^{d_2} - \aalpha^{d_1}\bigg] K\bigg\}
\left( 1 + \frac{\alpha_s(a^{-1})}{4\pi}D_L\right)\, ,
\label{eq:zlmatching}\end{eqnarray}
where $J$ and $K$ are constants obtained from the anomalous dimension
matrix of the operators $O_L$ and $O_S$ in the continuum effective 
theory, and $D_L$ is a constant (which arises in step (i) of the matching
procedure). The terms containing both the scales $m_b$ and $a^{-1}$ in
Eq.~(\ref{eq:zlmatching}) arise from the evolution in step (ii) and
the coefficients $C_L$ and $C_S$ from the matching in step (iii). 

To obtain $\bb$ we divide the result for $M(m_b)$ by $f_B^2$, and so
in the denominator there is a factor of $Z_A^2$, where $Z_A$ is the
constant relating the physical axial current to that in the lattice
static theory. $Z_A$ is also calculated in perturbation theory.

We consider three methods which have recently been used to treat the
perturbative corrections; these methods differ only at higher orders
in perturbation theory for which the coefficients are unknown. They
are therefore equivalent at NLO. Because some
of the coefficients in the $Z_i$'s 
are large, this leads to significant 
uncertainties in the result for $\bb$. 
The three methods are:
\begin{description}
\item[M1:] the expressions for each of the $Z_i$ are evaluated exactly
as in the example of Eq.~(\ref{eq:zlmatching}),
\item[M2:] the terms of $O(\alpha_s^2)$ in each $Z_i$ are 
dropped ($i=L,N,R,S,A$),
\item[M3:] the terms of $O(\alpha_s^2)$ in each $Z_i/Z_A^2$ are 
dropped ($i=L,N,R,S$).
\end{description}
M1 and M2 are the methods used by Gim\'enez and Martinelli
in~\cite{gm:static}, while M3 is the ``fully-linearised'' method of
the Kentucky group~\cite{ken:bb}.

For the numerical estimates in \tabref{bbstat} we adopt the same
choice of parameters as Gim\'enez and Martinelli in~\cite{gm:static},
using $\Lambda^{(4)}_{\overline\mathrm{MS}} = 200\mev$, $n_f=4$ and
$m_b = 5\gev$ for the lattice-to-continuum matching to find
$\bb(m_b)$~\footnote{In step (i) of the matching procedure we use
$\alpha_V(q^*a)$ evaluated at $q^* = 2.18 a^{-1}$ as the strong
coupling constant~\cite{lepage-mackenzie}.}.
To obtain $\rgbb^{\mathrm{nlo}}$ we set $n_f = 5$ in
\eqref{bbhat} and keep $\alpha_s$ continuous at the threshold at
$m_b$. Where necessary, we have taken the raw matrix elements from the
different groups and have simply propagated the errors in the linear
combination which determines $\bb$. 

As \tabref{bbstat} shows, because of the large coefficients, there are
significant differences between the formally equivalent procedures M1,
M2 and M3. Furthermore, the results from UKQCD~\cite{ukqcd:staticb}
and Gim\'enez and Martinelli~\cite{gm:static}, which use the same
action (i.e.\ the tree-level improved SW action corresponding to $\csw
= 1$) for the light quarks, agree when subjected to the same analysis
procedure, whereas the Kentucky group results~\cite{ken:bb}, using the
unimproved Wilson action ($\csw = 0$) but with tadpole-improved
perturbation theory in the matching, are significantly larger. A
similar conclusion is reached by Wittig
in~\cite{how:ijmp-review}. Because of these uncertainties, as
mentioned above, we will not make further use of these results, but
will use those obtained with propagating quarks to obtain the value of
$\bb$.

\subsection{$\bb$ obtained using propagating $b$-quarks}
\label{subsec:bbprop}

Calculations with propagating heavy quarks are reported in
\tabref{bbprop}. We have scaled all results to a common scale $m_b =
5\gev$, with the same parameter choice as for the static results
above, and then converted both to $\rgbb$ at NLO. The results show no
observable dependence on the lattice spacing, although the authors of
Ref.~\cite{bbs:lat96} perform a linear extrapolation to the continuum
limit, which is the reason for the relatively large error in the
corresponding result. Although the quoted errors are largely
statistical, the different groups treat the perturbative matching in
different ways, so that it is not appropriate to simply perform a weighted
average. Our preferred estimate based on the results in  
\tabref{bbprop} is
\begin{equation}
\rgbb^{\mathrm{nlo}} = 1.4(1)\ .
\label{eq:bbbest}
\end{equation}
In estimating the error we have assumed that the results are almost
independent of the lattice spacing and have not tried to quantify the
effects of quenching~\footnote{The second error in the result from
Ref.~\cite{soni:lat95} is the authors' estimate of the quenching
errors.}.  Calculations to be performed in the near future will use
improved actions and operators (reducing the discretization errors to
ones which are quadratic in the lattice spacing) and will avoid
lattice perturbation theory by using non-perturbative renormalization.

\begin{table}
\vspace{-8pt}
\caption[]{Results for the mixing parameter $\bb$ obtained in the
quenched approximation using propagating $b$-quarks. The authors'
results for $\bb(\mu)$ at scale $\mu$ or for $\bb(m_b)$ with $m_b =
5\gev$ are quoted. They are then scaled to $m_b$ and converted to
$\rgbb^{\mathrm{nlo}}$, the renormalization group invariant $B$
parameter. The results may differ slightly from those in the original
articles owing to the particular choice of parameters used here:
authors' numbers are shown in upright type, numbers produced by us in
oblique type.}
\label{tab:bbprop}
\kern1em
\begin{center}
\begin{tabular}{rccllll}
\hline
\ts  & Yr & $\beta$ & $\mu/\!\gev$ & $\bb(\mu)$ & $\bb(m_b)$
 & $\rgbb^{\mathrm{nlo}}$ \\[0.4ex]
\hline
\ts BBS~\cite{bbs:lat96} & 97 & $\infty$ & 2 & 1.02(13) & \sl0.96(12) &
  \sl1.53(19) \\
JLQCD~\cite{jlqcd:bb-lat95} & 96 & 6.3 & & & 0.840(60) & \sl1.34(10) \\
JLQCD~\cite{jlqcd:bb-lat95} & 96 & 6.1 & & & 0.895(47) & \sl1.42(7)  \\
BS~\cite{soni:lat95} & 96 & $\infty$ & 2 & 0.96(6)(4) & \sl0.90(6)(4) &
 \sl1.44(9)(6) \\
ELC~\cite{elc:fb92} & 92 & 6.4 & 3.7 & 0.86(5) & \sl0.84(5) &
  \sl1.34(8) \\
BDHS~\cite{bdhs} & 88 & 6.1 & 2 & 1.01(15) & \sl0.95(14) & \sl1.51(22)
 \\[0.4ex]
\hline
\end{tabular}
\end{center}
\end{table}

The relevant quantity for $B$--$\bar B$ mixing is $f_B^2
\rgbb$. Taking the result in \eqref{bbbest} above for
$\rgbb^{\mathrm{nlo}}$ with $f_B = 170(35)\mev$ from \eqref{fbbest}
gives
\begin{equation}
f_B\sqrt{\rlap{$\phantom{\hat B}$}\smash{\rgbb^{\mathrm{nlo}}}} = 201(42)\mev
\label{eq:xibbest}
\end{equation}
as our lattice estimate. 
An interesting dimensionless quantity is the ratio
\begin{equation}
\xi\equiv {\fbs \sqrt{\rlap{$\phantom{\hat B}$}\smash{\rgbbs}}  \over
 \fbd \sqrt{\rlap{$\phantom{\hat B}$}\smash{\rgbbd}}}
\end{equation}
For propagating quarks, combining the result $f_{B_s}/\!f_B = 1.14(8)$
from \eqref{fbs/fbbest} with $B_{B_s}/B_B =
1.00(3)$~\cite{bernardhf97} gives
\begin{equation}
\xi = 1.14(8).
\label{eq:xibest}\end{equation}
A recent direct extraction of the matrix element $M(\mu)$ gives the
ratio $r_{sd} \equiv \xi^2 m_{B_s}^2/m_{B_d}^2 =
1.54(13)(32)$~\cite{bbs:lat96}.  In the static case, Gim\'enez and
Martinelli~\cite{gm:static} find $r_{sd} = 1.43(7)$ by combining
$f_{B_s}/f_B = 1.17(3)$ and $B_{B_s}/B_B = 1.01(1)$, and $r_{sd} =
1.35(5)$ from a direct evaluation of the four-quark matrix element
measured on the same gauge configurations. The results of the two
methods are quite consistent, but future calculations should improve
on the precision of $r_{sd}$.

The above results are largely from quenched calculations. For
$B_{B_s}/B_B$, numerical evidence suggests a small increase on
two-flavour dynamical configurations~\cite{soni:lat95} but the chiral
loop estimate~\cite{sharpe-zhang,sharpe:lat96} is for a decrease of
$-0.04$ in the ratio. We must wait for reliable simulations with
dynamical quarks before the size of quenching effects can be
determined with confidence.

\section{Semileptonic Decays of $D$ and $B$-mesons}
\label{sec:sl}

\begin{figure}
\begin{center}
\begin{picture}(180,70)(20,10)
\SetWidth{2}\ArrowLine(10,41)(43,41)
\Text(15,35)[tl]{\boldmath$B$}
\SetWidth{0.5}
\Oval(100,41)(20,50)(0)
\SetWidth{2}\ArrowLine(157,41)(190,41)
\Text(180,35)[tl]{\boldmath$D,\,D^*,\,\pi,\,\rho$}
\SetWidth{0.5}
\Vertex(100,61){3}
\GCirc(50,41){7}{0.8}\GCirc(150,41){7}{0.8}
\Text(75,48)[b]{$b$}\Text(117,48)[b]{$c,u$}
\Text(100,16)[t]{$\bar q$}
\Text(100,71)[b]{$V{-}A$}
\ArrowLine(101,21)(99,21)
\ArrowLine(70,57)(72,58)\ArrowLine(128,57.5)(130,56.5)
\end{picture}
\caption{Diagram representing the semileptonic decay of the $B$-meson.
$\bar q$ represents the light valence antiquark, and the black circle
represents the insertion of the $V$--$A$ current with the appropriate
flavour quantum numbers.}
\label{fig:sl}
\end{center}
\end{figure}
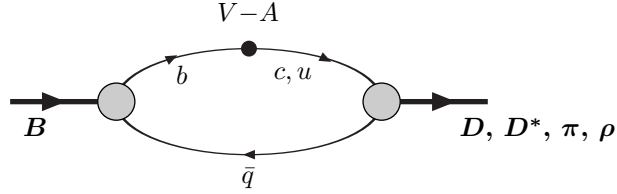

We now discuss semileptonic decays of $D$ and $B$-mesons, considering
in turn the cases in which the $c$-quark decays into an $s$- or
$d$-quark and the $b$-quark decays into a $c$- or $u$-quark. The
$B$-meson decay is represented in~\figref{sl}, though the diagram
could simply be relabelled to describe the $D$-meson decay. It is
convenient to use space-time symmetries to express the matrix elements
in terms of invariant form factors (using the helicity basis for these
as defined below).  When the final state is a pseudoscalar meson $P$,
parity implies that only the vector component of the $V{-}A$ weak
current contributes to the decay, and there are two independent form
factors, $f^+$ and $f^0$, defined by
\begin{eqnarray}
\langle P(k)| V^\mu|B(p)\rangle & = &
  f^+(q^2)\left[(p+k)^\mu -
  \frac{m_B^2 - m_P^2}{q^2}\,q^\mu\right] \nonumber\\ 
& & \mbox{} + f^0(q^2)\,\frac{m_B^2 - m_P^2}{q^2}\,q^\mu\ ,
\label{eq:ffpdef}
\end{eqnarray}
where $q$ is the momentum transfer, $q=p{-}k$, and $B(p)$ denotes
either a $B$ or $D$ meson. When the final-state hadron is a vector
meson $V$, there are four independent form factors:
\begin{eqnarray}
\langle V(k,\varepsilon)| V^\mu|B(p)\rangle & = &
  \frac{2V(q^2)}{m_B+m_V}\epsilon^{\mu\gamma\delta\beta}
  \varepsilon^*_\beta p_\gamma k_\delta \label{eq:ffvvdef}\\ 
\langle V(k,\varepsilon)| A^\mu|B(p)\rangle  & = &  
  i (m_B {+} m_V) A_1(q^2) \varepsilon^{*\,\mu}
 - 
  i\frac{A_2(q^2)}{m_B{+}m_V} \varepsilon^*\!\dotprod p\,
  (p {+} k)^\mu \nonumber \\
& & \mbox{} + i \frac{A(q^2)}{q^2} 2 m_V 
  \varepsilon^*\!\dotprod p\, q^\mu\ , \label{eq:ffvadef}
\end{eqnarray}
where $\varepsilon$ is the polarization vector of the final-state
meson, and $q = p{-}k$.  Below we shall also discuss the form factor
$A_0$, which is given in terms of those defined above by $A_0 = A +
A_3$, with
\begin{equation}
A_3 = \frac{m_B + m_V}{2 m_V}A_1 - 
\frac{m_B - m_V}{2 m_V}A_2\ .
\label{eq:a3def}
\end{equation} 

\subsection{Semileptonic $D$ Decays}
\label{subsec:slDdecay}

The decays $D\to K l^+\nu_l$ and $D\to K^* l^+ \nu_l$ provide a good
test for lattice calculations since the relevant CKM matrix element
$V_{cs}$ is well constrained in the standard model. The form factors
for the decays $D\to \pi l^+\nu_l$ and $D\to\rho l^+\nu_l$ can also be
computed in lattice simulations. As explained in the introduction,
charm quarks are light enough to be simulated directly (though one
still needs to be wary of mass-dependent discretization
errors). Furthermore, strange quarks can also be simulated directly,
so for $D\to K$ or $K^*$ decays there is only one quark for which a
chiral extrapolation needs to be performed. For semileptonic $D$-meson
decays the whole physical phase space can be sampled~\footnote{In
addition, one obtains the form factors for unphysical, negative,
values of $q^2$.}, while keeping the spatial momenta of the initial
and final state mesons small in order to minimise the
momentum-dependent discretization errors.

\begin{figure}
\hbox to\hsize{\hss\epsfxsize=0.6\hsize
\epsffile[38 47 510 503]{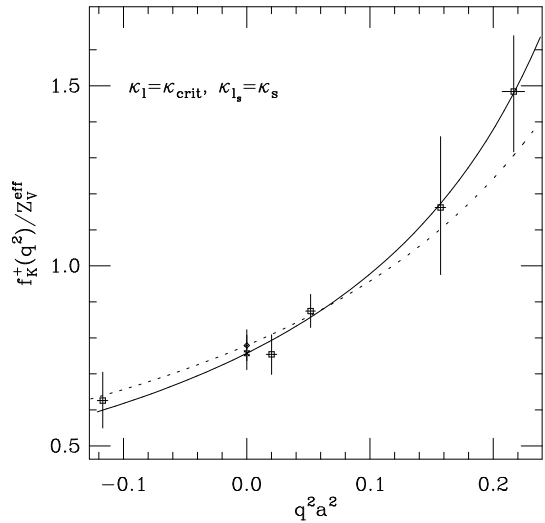}\hss}
\caption[]{UKQCD lattice results for the form factor $f^+(q^2)$ in
semileptonic $D\to K$ decay~\cite{ukqcd:d-semilept}. The form factor
is plotted as a function of the dimensionless variable $q^2a^2$ and
before multiplication by the appropriate lattice-to-continuum
normalization factor $Z_V^{\mathrm{eff}}$. The label $\kappa_l=\kcrit,
\kappa_{l_s}=\kappa_s$ shows that the light quark masses have been
extrapolated to give the physical $D\to K$ form factor. The curves
show fits of pole dominance behaviour to the data: the solid line is a
two parameter fit, the dashed line a fit where the pole mass is fixed
to the corresponding $1^-$ vector meson mass determined in the same
lattice calculation. The cross and diamond show the interpolations of
the fits to $q^2=0$.}
\label{fig:ukqcd-dtok-fplus}
\end{figure}

Although the lattice calculations actually measure the $q^2$
dependence of the form factors, as shown by the example of $f^+$ for
the $D\to K$ decay in
\figref{ukqcd-dtok-fplus}~\cite{ukqcd:d-semilept}, we follow the
standard practice of quoting values at $q^2=0$. In contrast to the
case for $B$ decays to be discussed below, we emphasise that this
involves an interpolation and so is relatively well controlled. We
refer the reader to the original papers for detailed discussions of
the $q^2$-dependence of the form factors.

\begin{figure}
\unit=0.88\hsize
\epsfxsize=\unit
\vspace*{0.06\unit}
\hbox to\hsize{%
\raisebox{0.33\unit}{\parbox[t][0.305\unit][s]{0.105\hsize}{\small\raggedleft
LMMS\par\vfill
BKS\par\vfill
ELC\par\vfill
APE\par\vfill
UKQCD\par\vfill
LANL\par\vfill
WUP\par\vfill
Expt}}\hfill
\vbox{\offinterlineskip
\epsffile{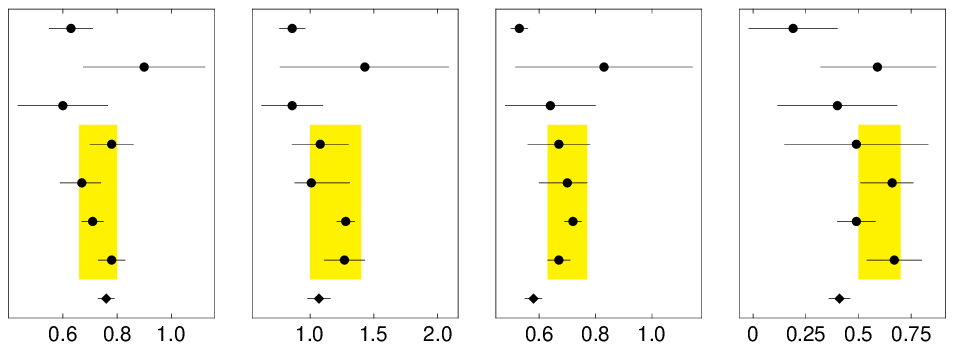}
\point 0.06 0.408 {f^{K,+}(0)}
\point 0.319 0.41 {V^{K^*}(0)}
\point 0.573 0.41 {A^{K^*}_1(0)}
\point 0.832 0.41 {A^{K^*}_2(0)}
}}
\caption[]{Lattice results for $D\to K$ and $D\to K^*$ semileptonic
decay form factors at zero momentum transfer, and comparison with
experimental results form the survey in Ref.~\cite{ryd:hf7}. The
shaded bands show our summary values, indicating which results they
are based on, as discussed in the text.}
\label{fig:dtokff}
\end{figure}

Lattice results for the $D\to K^{(*)}$ form factors are collected in
\tabref{dtok-lattice} and illustrated in \figref{dtokff}, while
results for the $D\to \pi,\rho$ form factors appear in
\tabref{dtopirho-lattice}. These are all from quenched simulations and
no group has performed a continuum extrapolation. Our summary view of
these results is obtained by considering the more recent results from
WUP~\cite{wup:hl-semilept},
LANL~\cite{lanl:semilept-lat94,lanl:semilept-lat95},
UKQCD~\cite{ukqcd:d-semilept} and APE~\cite{ape:hl-semilept} which all
use either the improved SW action or the Wilson action with a KLM
normalization. The summary values are presented in
\tabref{slDdecayresults}, along with recent experimental world
averages~\cite{ryd:hf7} in the $D\to K,K^*$ case. The values quoted
reflect the fact that $f^+$ and $A_1$ are the best measured while the
$D\to \pi,\rho$ form factors are smaller with slightly larger errors.
The shaded bands in \figref{dtokff} show our estimates in the $D\to
K,K^*$ case and indicate which results they are based on.

One sees that the lattice and experimental results agree rather well.
The lattice values for $A_1$ and $A_2$ are both high compared to
experiment: however, these depend on the correct normalization of the
lattice axial vector current which is less well known than the vector
current normalization needed for $f^+$ and $V$. In particular, the
non-degeneracy of the $c$- and $s$-quarks means that there is no
natural normalization condition to use for the weak current. This
contrasts with the situation for heavy-to-heavy semileptonic decays,
described in \subsecref{vcb}, where one can make use of the
conservation of the vector current of degenerate quarks.

\begin{table}
\vspace{-8pt}
\caption[]{Collected lattice results for $D\to K, K^*$ semileptonic
decay form factors at $q^2=0$.}
\label{tab:dtok-lattice}
\kern1em
\begin{center}
\begin{tabular}{rllllll}
\hline
\ts & Yr & $f^+(0)$ &$f^0(0)$ & $V(0)$ & $V/A_1$ \\[0.4ex]
\hline
\ts
WUP~\cite{wup:hl-semilept} & 97 & $0.78(5)$ &$0.78(5)$ &
 $1.27(16)$ &\\
LANL~\cite{lanl:semilept-lat94,lanl:semilept-lat95} & 95--96
  & $0.71(4)$ &$0.73(3)$ & $1.28(7)$ & $1.78(7)$\\
UKQCD~\cite{ukqcd:d-semilept} & 95 & 0.67\errp78 & $0.65(7)$ &
  1.01\errp{30}{13} & 1.4\errp{5}{2}  \\ 
APE~\cite{ape:hl-semilept} & 95 & $0.78(8)$ & & $ 1.08(22)$ & $1.6(3)$ \\
ELC~\cite{elc:hl-semilept} & 94 & 0.60(15)(7)  & & $0.86(24)$& $1.3(2)$ \\
BKS~\cite{bks} & 91--92 & $0.90(8)(21)$ & $0.70(8)(24)$ 
&$1.43(45)\errp{48}{49}$ & $1.99(22)\errp{31}{35}$ \\
LMMS~\cite{lmms} & 89--92 & $0.63(8)$ & & $ 0.86(10)$ & $1.6(2)$\\[0.4ex]
\hline\\[1ex]\hline
\ts & Yr & $A_1(0)$ & $A_2(0)$ & $A_2/A_1$ & $A_0(0)$ \\[0.4ex]
\hline
\ts
WUP~\cite{wup:hl-semilept} & 97 & $0.67(4)$ &$0.67(13)$ & &\\
LANL~\cite{lanl:semilept-lat94,lanl:semilept-lat95} & 95--96
  & $0.72(3)$ &$0.49(9)$ & $0.68(11)$ & $0.84(3)$ \\
UKQCD~\cite{ukqcd:d-semilept} & 95 & 0.70\errp7{10} & 0.66\errp{10}{15} &
 $0.9(2)$  & 0.75\errp5{11}  \\ 
APE~\cite{ape:hl-semilept} & 95 & $0.67(11)$ &$0.49(34)$ &$0.7(4)$ & \\
ELC~\cite{elc:hl-semilept} & 94 & $0.64(16)$ & $0.40(28)(4)$ & $0.6(3)$ &  \\
BKS~\cite{bks} & 91--92 & $0.83(14)(28)$ &$0.59(14)\errp{24}{23}$ &
 $0.70(16)\errp{20}{15}$& $0.71(16)(25)$ \\
LMMS~\cite{lmms} & 89--92 & $0.53(3)$ & $0.19(21)$ & $ 0.4(4)$ & \\[0.4ex]
\hline
\end{tabular}
\end{center}
\end{table}

\begin{table}
\vspace{-8pt}
\caption[]{Collected lattice results for $D\to \pi, \rho$ semileptonic
decay form factors at $q^2=0$.}
\label{tab:dtopirho-lattice}
\kern1em
\begin{center}
\begin{tabular}{rllllll}
\hline
\ts  & Yr & $f^+(0)$ &$f^0(0)$ & $V(0)$ & $V/A_1$ \\[0.4ex]
\hline
\ts
WUP~\cite{wup:hl-semilept} & 97 & $0.73(6)$ &$0.73(6)$ &
 $1.18(17)$ &\\
LANL~\cite{lanl:semilept-lat94} & 95
  & $0.56(8)$ &$0.62(5)$ & $1.18(15)$ & $1.77(16)$\\
UKQCD~\cite{ukqcd:d-semilept} & 95 & 0.61\errp{12}{11} &
 $0.53\errp{12}{11}$ & 0.95\errp{29}{14} &  \\ 
BKS~\cite{bks} & 91--92 & $0.84(12)(35)$ & $0.62(6)(34)$ 
&$1.07(49)(35)$ & $2.01(40)(32)$ \\
LMMS~\cite{lmms} & 89--92 & $0.58(9)$ & & $0.78(12)$ \\[0.4ex]
\hline\\[1ex]\hline
\ts & Yr & $A_1(0)$ & $A_2(0)$ & $A_2/A_1$ & $A_0(0)$ \\[0.4ex]
\hline
\ts
WUP~\cite{wup:hl-semilept} & 97 & $0.63(5)$ &$0.64(14)$ & &\\
LANL~\cite{lanl:semilept-lat94} & 95
  & $0.67(7)$ &$0.44(24)$ & $0.67(31)$ \\
UKQCD~\cite{ukqcd:d-semilept} & 95 & 0.63\errp69 & 0.51\errp{10}{15} &
 & 0.70\errp5{12}  \\ 
BKS~\cite{bks} & 91--92 & $0.65(15)\errp{24}{23}$ &$0.59(31)\errp{28}{25}$ &
 $0.89(37)\errp{22}{19}$& $0.64(17)(21)$ \\
LMMS~\cite{lmms} & 89--92 & $0.45(4)$ & $0.02(26)$ \\[0.4ex]
\hline
\end{tabular}
\end{center}
\end{table}

\begin{table}
\vspace{-8pt}
\caption[]{Summary of lattice and experimental results for $D\to K,
K^*$ and $D\to \pi,\rho$ semileptonic decay form factors at
$q^2=0$. The experimental numbers are taken from the survey
in Ref.~\cite{ryd:hf7}.}
\label{tab:slDdecayresults}
\kern1em
\begin{center}
\begin{tabular}{l@{\qquad}ll@{\qquad}l}
\hline
\ts & \multicolumn{2}{@{}l}{$D\to K,K^*$} & $D\to\pi,\rho$ \\
 & lattice & expt & lattice \\
\hline \ts
$f^+(0)$ & 0.73(7) & 0.76(3) & 0.65(10) \\
$V(0)$   & 1.2(2)  & 1.07(9) & 1.1(2) \\
$A_1(0)$ & 0.70(7) & 0.58(3) & 0.65(7) \\
$A_2(0)$ & 0.6(1)  & 0.41(5) & 0.55(10) \\[0.4ex]
\hline
\end{tabular}
\end{center}
\end{table}

\subsection{Semileptonic $B\to D$ and $B\to D^*$ Decays}
\label{subsec:vcb}

Semileptonic $B\to D^*$ and, more recently, $B \to D$ decays are used
to determine the $V_{cb}$ element of the CKM matrix. Heavy quark
symmetry is rather powerful in controlling the theoretical description
of these heavy-to-heavy quark transitions, as described by Neubert in
the chapter on ``$B$ Decays and the Heavy Quark Expansion'' in this
volume~\cite{neubert}. In describing lattice results for these decays,
we will quote theoretical results and direct the reader
to Ref.~\cite{neubert} for details and references.

In the heavy quark limit all six form factors in
Eqs.~(\ref{eq:ffpdef}--\ref{eq:ffvadef}) are related and there is just
one universal form factor $\xi(\omega)$, known as the Isgur--Wise (IW)
function which contains all the non-perturbative QCD
effects. Specifically:
\begin{eqnarray}
f^+(q^2) & = & V(q^2) = A_0(q^2) = A_2(q^2) 
\nonumber\\ 
& = & \left[1 - 
\frac{q^2}{(m_B + m_D)^2}\right]^{-1} A_1(q^2) = \frac{m_B+m_D}
{2\sqrt{m_Bm_D}}\,\xi(\omega)\ ,
\label{eq:iw}
\end{eqnarray}
where $\omega = v_B\dotprod v_D$ is the velocity transfer
variable. Here the label $D$ represents the $D$- or $D^*$-meson as
appropriate (pseudoscalar and vector mesons are degenerate in this
leading approximation). Vector current conservation implies that the
IW-function is normalized at zero recoil, i.e.\ that $\xi(1) =1$. This
property is particularly important in the extraction of $V_{cb}$.

The relations in Eq.~(\ref{eq:iw}) are valid up to perturbative and
power corrections. Allowing for corrections to the heavy quark limit,
one writes the decay distribution for $B\to D^*$ as
\begin{eqnarray}
\frac{d\Gamma}{d\omega} &=& 
 \frac{G_F^2}{48\pi^3} (m_B{-}m_{D^*})^2 m_{D^*}^3
    \sqrt{\omega^2{-}1}\,(\omega{+}1)^2 \nonumber \\
& &  \mbox{}\times \left[ 1 + \frac{4\omega}{\omega{+}1} 
     \frac{m_B^2-2\omega m_Bm_{D^*}+m_{D^*}^2}{(m_B-m_{D^*})^2}
    \right]
    |V_{cb}|^2\, {\cal F}^2(\omega)\ ,
\label{eq:distr}
\end{eqnarray}
where ${\cal F}(\omega)$ is the ``physical form factor'' given by the
IW-function combined with perturbative and power corrections. To
extract $|V_{cb}|$ one extrapolates measurements of the product
$\mathcal{F}(\omega) |V_{cb}|$ to the zero-recoil point, $\omega=1$,
and uses a theoretical evaluation of the normalization
$\mathcal{F}(1)$~\cite{neubert}.

A theoretical understanding of the shape of the physical form factor
would be useful to guide the extrapolation of the experimental data,
which currently show rather a wide variation for the slope and 
intercept~\cite{neubert,lkg:ichep96},
and also as a test of our understanding of the QCD effects. We expand
${\cal F}$ as a power series in $\omega -1$:
\begin{equation}
{\cal F}(\omega) = {\cal F}(1)\, \left[1 - \hat\rho^2\,(\omega -1)
+\hat c\, (\omega -1 )^2 + \cdots\right].
\label{eq:ftaylor}
\end{equation}
The slope parameter $\hat\rho^2$ differs from the slope parameter
$\rho^2$ of the IW function itself by heavy quark symmetry violating
corrections~\cite{neubert},
\begin{equation}
\hat\rho^2 = \rho^2 + (0.16\pm 0.02) + \mbox{power corrections}.
\label{eq:rhohat}
\end{equation}

To discuss lattice results for the shape of the IW function, and to
search for corrections to the relations obtained using heavy quark
symmetry, it is convenient to work with the following set of form
factors (expressed in terms of four-velocities) which in the heavy
quark limit either vanish or are equal to the IW function: 
\begin{eqnarray*}
{\bra{D(v')} \bar c\gamma^\mu b \ket{B(v)} \over
 \sqrt{m_B m_D}} &\!\!=\!\!& (v{+}v')^\mu h_+(\w) +
    (v{-}v')^\mu h_-(\w) \\
{\bra{D^*(v')} \bar c\gamma^\mu b \ket{B(v)} \over
 \sqrt{m_B m_D}} &\!\!=\!\!& i \epsilon^{\mu\nu\alpha\beta} \epsilon^*_\nu
 v'_\alpha v_\beta h_V(\w) \\
{\bra{D^*(v')} \bar c\gamma^\mu \gamma_5 b \ket{B(v)} \over
 \sqrt{m_B m_D}} &\!\!=\!\!& (\w{+}1)\epsilon^{*\mu} h_{A_1}(\w)
   - \epsilon^*\!\dotprod v \big(v^\mu h_{A_2}(\w) +
     v'^\mu h_{A_3}(\w)\big)
\end{eqnarray*}
where
\begin{equation}
h_i(\w) = \big(\alpha_i + \beta_i(\w) + \gamma_i(\w)\big) \xi(\w)
\end{equation}
with
\begin{equation}
\alpha_+ = \alpha_V = \alpha_{A_1} = \alpha_{A_3} = 1, \qquad
 \alpha_- = \alpha_{A_2} = 0.
\end{equation}
The $\beta_i$ and $\gamma_i$ denote perturbative and power corrections
(in $1/m_{b,c}$) respectively. The statement of Luke's
theorem~\cite{luke} is
\begin{equation}
\gamma_{+,A_1}(1) = O(\lqcd^2/m_{c,b}^2).
\end{equation}

The principal difficulty for lattice calculations is to separate the
physical heavy quark mass dependence due to power corrections from the
unphysical one due to mass-dependent discretization errors. One must
also address the question of lattice-to-continuum matching. We will
illustrate our discussion using the analysis procedure applied by the
UKQCD collaboration~\cite{ukqcd:iw} $\big($who use the SW action for
all quarks so that the leading mass-dependent discretization errors
are formally reduced to $O(\alpha_s m_Qa)$ and $O(m_Q^2
a^2)$$\big)$. Consider the case of $h_+$. For this form factor we have
the protection of Luke's theorem at zero recoil and, for degenerate
($Q= Q'$) transitions, conservation of the vector current $\bar
Q\gamma_\mu Q$ provides the further constraints:
\begin{equation}
\beta_+(1;m_Q,m_Q) = 0, \qquad \gamma_+(1;m_Q,m_Q) = 0.
\end{equation}
The correct normalization of the vector current can be assured by
requiring that the electric charge of the meson be 1. 
We therefore define the continuum form factor by,
\begin{equation}
h_+(\w;m_Q,m_{Q'}) \equiv
  \big(1 + \beta_+(1;m_Q,m_{Q'})\big)
 {h_+^{\mathrm{L}}(\w;m_Q,m_{Q'})\over h_+^{\mathrm{L}}(1;m_Q,m_{Q'})}\ ,
\label{eq:hplusdef}
\end{equation}
where $h_i^{\mathrm{L}}(\w;m_Q,m_{Q'})$ is the (un-normalized) form
factor calculated directly in the lattice simulation. This definition
partially removes discretization errors and also removes
$\w$-independent power corrections while maintaining the known
normalization conditions. If the remaining power corrections are
small, then
\begin{equation}
h_+(\w)\over 1 + \beta_+(\w)
\end{equation}
is effectively the IW function, $\xi(\w)$. This is convenient for
extracting $\xi(\w)$, but the definition of \eqref{hplusdef} precludes
a determination of the zero-recoil power corrections. These
corrections should be small, being suppressed by two powers
of the heavy quark mass. However, applying an analogous procedure to
the $h_{A_1}(\w)$ form factor relevant for $B\to D^*$ decays will not
allow the $1/m_c^2$ corrections to $\mathcal{F}(1)$, one of the
dominant theoretical uncertainties, to be determined.

\begin{table}
\vspace{-8pt}
\caption{Values of the Slope of the IW--function of a heavy meson, obtained
using lattice QCD. The table indicates which functional form from
\eqref{iwforms} has been fitted to and which form factor has been used
in the extraction. The systematic error in the UKQCD results
incorporates the variation from fitting to all functional forms in
\eqref{iwforms}. BSS note that $\rho_{u,d}^2$ is 12\% smaller than
$\rho_s^2$, but do not quote a separate result.}
\label{tab:rhosq}
\kern1em
\centering
\begin{tabular}{rlllll}
\hline
\ts & Yr & $\rho^2_{u,d}$ & $\rho^2_s$ & fit  & using \\[0.4ex]
\hline
\ts
LANL~\cite{lanl:semilept-lat95} & 96 & 0.97(6) & & NR & $h_+$ \\
UKQCD~\cite{ukqcd:iw} & 95 & $0.9\errp23\errp42$ & $1.2\errp22\errp21$
 & NR & $h_+$ \\
UKQCD~\cite{lpl:zako} & 94 & $0.9\errp45\errp91$ & $1.2\errp33\errp71$
 & NR & $h_{A_1}$ \\
BSS~\cite{bss:iw} & 93  & & 1.24(26)(36) & lin & $h_+$ \\
BSS~\cite{bss:iw} & 93  & & 1.41(19)(41) & NR  & $h_+$ \\[0.4ex]
\hline
\end{tabular}
\end{table}

UKQCD confirm~\cite{ukqcd:iw} that their results for $h_+/(1+\beta_+)$
are indeed independent of the heavy quark masses and hence demonstrate,
within the available precision,
that there \emph{is} an IW function. Moreover, a similar analysis for
the $h_{A_1}$ form factor reveals the same function~\cite{lpl:zako},
so the IW function appears indeed to be universal. Performing
extrapolations to the light ($u$-, $d$-) and strange ($s$-) quark
masses and fitting to
\begin{equation}
\xi(\w) = \cases{\xi_{\mathrm{NR}}(\w) \equiv {2\over\w{+}1}
  \exp\big(-2(\rho^2{-}1){\w{-}1\over\w{+}1}\big)& NR\cr
  1 - \rho^2(\w-1)& linear\cr
  1 - \rho^2(\w-1) + {c\over2}(\w-1)^2& quadratic\cr}
\label{eq:iwforms}
\end{equation}
gives lattice determinations of the slope of the IW function, as
listed in \tabref{rhosq}. Since `the' IW function is different for
different light degrees of freedom, the results in the table are
labelled with subscripts $u,d$ or $s$ as appropriate.

\begin{figure}
\hbox to\hsize{\hss\epsfxsize=0.6\hsize\epsffile{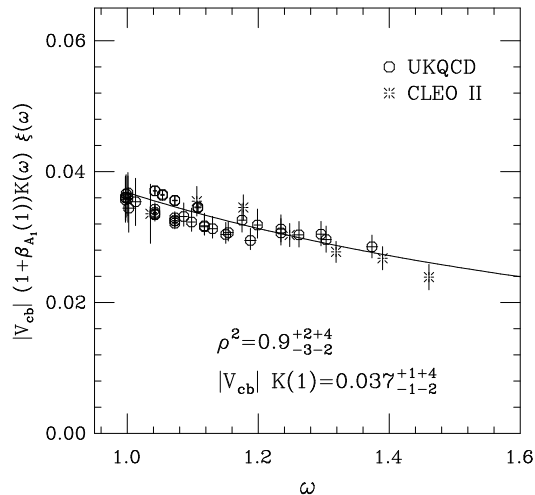}\hss}
\caption[]{Fit of the UKQCD lattice results for $|V_{cb}|{\cal
F}(\omega)$~\cite{ukqcd:iw} to the experimental data from the CLEO
collaboration~\cite{cleoii}.}
\label{fig:iw}
\end{figure}

In \figref{iw} we show the comparison of the UKQCD lattice
results~\cite{ukqcd:iw} with $B\to D^*$ data (in 1995) from the CLEO
collaboration~\cite{cleoii}. A chi-squared fit is made to the
experimental data for $|V_{cb}|[1 + \beta_{A_1}(1)]K(\w)\xi_{u,d}(\w)
\equiv |V_{cb}| \mathcal{F}(\w)$. $K(\w)$ incorporates radiative
corrections away from zero-recoil $\big($$\beta_{A_1}(\w)$$\big)$,
nonperturbative power corrections $\big($$\gamma_{A_1}(\w)$$\big)$ and
the contributions of the other form factors to the rate. The current
lattice calculations cannot distinguish the $\w$ dependence of
$K(\w)$, so it is taken to be a constant and hence $\rho_{u,d}^2$ and
$\hat\rho^2$ are not distinguished. The slope of the IW function is
constrained to the lattice result in the fit so that the only free
parameter is $|V_{cb}|\mathcal{F}(1)$. The result of the fit
is~\cite{ukqcd:iw}
\begin{equation}
|V_{cb}| \mathcal{F}(1) = 0.037\err11\err22\err41.
\end{equation}

We should also mention $B\to D$ semileptonic decays, which are
beginning to be measured experimentally~\cite{artuso,lkg:ichep96} with
good precision, despite the helicity suppression in $d\Gamma(B\to D
l\bar\nu_l)/d\w$. The differential decay rate depends on both
$h_+(\w)$ and $h_-(\w)$. However, $h_-$ is rather poorly determined to
date in lattice calculations, so that it is difficult to evaluate the 
$O(1/m_Q)$ corrections.

Direct lattice calculations of the IW function are being undertaken
using discretizations of the heavy quark effective
theory~\cite{man-og}$^-$\cite{milc:iwfn-lat97}. These studies
are very interesting, but the results are not yet useful for
phenomenology. An interesting theoretical feature of this approach is
the formulation of the HQET at non-zero velocity in Euclidean
space~\cite{man-og1}$^-$\cite{aglietti}.

Finally we note that a first lattice study of the semileptonic decays
$\Lambda_b \to \Lambda_c l \nu$ and $\Xi_b\to \Xi_c l\nu$ has recently
been performed~\cite{ukqcd:baryonff}, giving predictions for the decay
distributions and the baryonic Isgur-Wise function. We refer the
interested reader to Ref.~\cite{ukqcd:baryonff} for details.

\subsection{Semileptonic $B\to \rho$ and $B\to \pi$ Decays
and the Rare Decay $\btokstargamma$}
\label{subsec:vub}

In this subsection we consider the heavy-to-light semileptonic decays
$B\to\rho$ and $B\to\pi$ which are now being used experimentally to
determine the $V_{ub}$ matrix element~\cite{lkg:ichep96,jrp:ichep96}.
Several groups have evaluated form factors for these decays using
lattice simulations~\cite{elc:hl-semilept,ape:hl-semilept,wup:hl-semilept,%
ukqcd:hlff}$^-$\cite{ukqcd:hlfits}
(see the recent review in~\cite{onogi:lat97}). We will also consider
the rare radiative decay $\btokstargamma$ which is related by heavy
quark and light flavour symmetries to the $B\to \rho$ semileptonic
decay, and which was observed experimentally for the first time
in 1993~\cite{cleo:BKstarGamma,cleo:rkstar}.

Form factors for semileptonic $B\to\pi$ decays were defined above in
\eqref{ffpdef} and for $B\to\rho$ decays in \eqref{ffvadef}. 
For completeness, we define here form factors for the
matrix element of the magnetic moment operator responsible for the
short distance contribution to the $\btokstargamma$ decay:
\begin{equation}
\langle K^*(k,\varepsilon) | \bar{s} \sigma_{\mu\nu} q^\nu b_R
 | B(p) \rangle
  =  \sum_{i=1}^3 C^i_\mu T_i(q^2),
\label{eq:bkstarME}
\end{equation}
where $q=p{-}k$, $\varepsilon$ is the polarization vector of the $K^*$ and
\begin{eqnarray}
C^{1}_\mu & = &
  2 \epsilon_{\mu\nu\lambda\rho} \varepsilon^{*\,\nu} p^\lambda k^\rho, \\
C^{2}_\mu & = &
  \varepsilon^*_\mu(m_B^2 - m_{K^*}^2) - \varepsilon\dotprod q (p+k)_\mu, \\
C^{3}_\mu & = & \varepsilon^*\!\dotprod q
  \left( q_\mu - \frac{q^2}{m_B^2-m_{K^*}^2} (p+k)_\mu \right).
\end{eqnarray}
$T_3$ does not contribute to the physical $\btokstargamma$ amplitude
for which $q^2=0$, and $T_1(0)$ and $T_2(0)$ are related by,
\begin{equation}
T_1(q^2{=}0) = i T_2(q^2{=}0).
\end{equation}
Hence, for the process $\btokstargamma$, we need to determine $T_1$
and/or $T_2$ at the on-shell point $q^2{=}0$.

Heavy quark symmetry is less predictive for heavy-to-light decays than
for heavy-to-heavy ones.  In particular, there is no normalization
condition at zero recoil corresponding to the condition $\xi(1)=1$,
which is so useful in the extraction of $V_{cb}$. The lack of such a
condition puts a premium on the results from nonperturbative
calculational techniques, such as lattice QCD. Heavy quark symmetry
does, however, give useful scaling laws for the behaviour of the form
factors with the mass of the heavy quark at fixed $\omega$. Moreover,
the heavy quark spin symmetry relates the $B\to V$ matrix
elements~\cite{iw:hqet,gmm} (where $V$ is a light vector particle) of
the weak current and magnetic moment operators, thereby relating the
amplitudes for the two processes $\btorho$ and $\btokstargamma$, up to
$SU(3)$ flavour symmetry breaking effects.

\begin{table}
\vspace{-8pt}
\caption[]{Leading $M$ dependence of form factors for heavy-to-light
decays in the helicity basis. The dependence follows from heavy quark
symmetry applied at \emph{fixed} velocity transfer~$\w$. Note that
only three of the four $A_i$ form factors for $B\to\rho l\nu$ are
independent.}
\label{tab:leadingM}
\kern0.5em
\hbox to\hsize{\hss
\begin{tabular}[t]{lll}
\hline
\ts form   & $t$-channel & leading $M$ \\[-0.3ex]
factor & exchange    & dependence \\
\hline
\multicolumn{3}{c}{\ts$B\to\rho l\nu$}\\
$V$ & $1^-$ & $M^{1/2}$ \\
$A_1$ & $1^+$ & $M^{-1/2}$ \\
$A_2$ & $1^+$ & $M^{1/2}$ \\
$A_3$ & $1^+$ & $M^{3/2}$ \\
$A_0$ & $0^-$ & $M^{1/2}$ \\
\hline
\end{tabular}
\hss
\begin{tabular}[t]{lll}
\hline
\ts form   & $t$-channel & leading $M$ \\[-0.3ex]
factor & exchange    & dependence \\
\hline
\multicolumn{3}{c}{\ts$B\to\pi l\nu$}\\
$f^+$ & $1^-$ & $M^{1/2}$ \\
$f^0$ & $0^+$ & $M^{-1/2}$ \\
\multicolumn{3}{c}{$B\to K^*\gamma$}\\
$T_1$ & $1^-$ & $M^{1/2}$ \\
$T_2$ & $1^+$ & $M^{-1/2}$\\
\hline
\end{tabular}
\hss}
\end{table}

\begin{figure}
\unit=\hsize
\hbox to\hsize{\hss\vbox{\offinterlineskip
\epsfxsize=\unit\epsffile[-2 99 280 186]{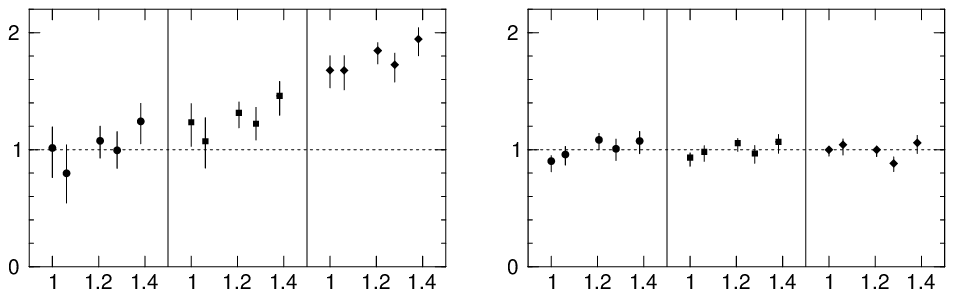}
\point 0.0 0.25 {\displaystyle{V\over2T_1}}
\point 0.5 0.25 {\displaystyle{A_1\over2iT_2}}
\point 0.125 0.095 \infty
\point 0.265 0.1 B
\point 0.405 0.1 D
\point 0.635 0.095 \infty
\point 0.775 0.1 B
\point 0.914 0.1 D
}\hss}
\kern-0.5ex
\hbox to\hsize{\hss
\hbox to\unit{\kern0.275\unit$\w$\kern0.49\unit$\w$\hfill}
\hss}
\caption[]{Ratios $V/2T_1$ and $A_1/2iT_2$ for five values of $\w$ at
three different heavy-light pseudoscalar masses, around the $D$ mass,
around the $B$ mass and in the infinite mass limit. The horizontal
dashed line denotes the heavy quark symmetry prediction in the
infinite mass limit. Data from Ref.~\cite{ukqcd:btorho}.}
\label{fig:hqs-test}
\end{figure}

For fixed $\w$ the scaling laws for the form factors given by heavy
quark symmetry are as follows:
\begin{equation}
f\big(q^2(\w)\big)\big|_{\w\mathrm{\ fixed}} \Theta = M^{\nu_f} \gamma_f
  \left(1+\frac{\delta_f}{M}+\frac{\epsilon_f}{M^2}+\cdots\right)
\label{eq:scaling}
\end{equation}
where $f$ labels the form factor, $M$ is the mass of the heavy-light
meson and $\Theta$ is a calculable leading logarithmic correction. The
leading $M$ dependences, $M^{\nu_f}$, are listed in
\tabref{leadingM}. Lattice calculations with propagating quarks use a
range of quark masses around the charm mass and generally employ these
scaling relations to extrapolate to the $B$ mass: this is the case for
results from ELC, APE and UKQCD. In the limit $M\to\infty$ we also
have the relations
\begin{equation}
A_1\big(q^2(\w)\big)=2iT_2\big(q^2(\w)\big),\quad
 V\big(q^2(\w)\big)=2T_1\big(q^2(\w)\big),
\label{eq:a1t2vt1hqs}
\end{equation}
at fixed $\w$. The UKQCD collaboration have checked the validity of
the relations in \eqref{a1t2vt1hqs}~\cite{ukqcd:btorho}, finding that
they are well satisfied in the infinite mass limit as shown in
\figref{hqs-test}. However, the ratio $V/2T_1$ already shows
significant deviations from the limiting value, 1, at the $B$
mass.

There are also kinematic constraints on the form factors at
$q^2=0$:
\begin{equation}
f^+(0) = f^0(0), \qquad T_1(0) = iT_2(0), \qquad A_0(0) = A_3(0),
\label{eq:ffkinconstraints}
\end{equation}
which will be useful below.

\begin{figure}
\unit0.7\hsize
\hbox to\hsize{\hss\vbox{\offinterlineskip
\epsfxsize\unit
\epsffile[-25 54 288 232]{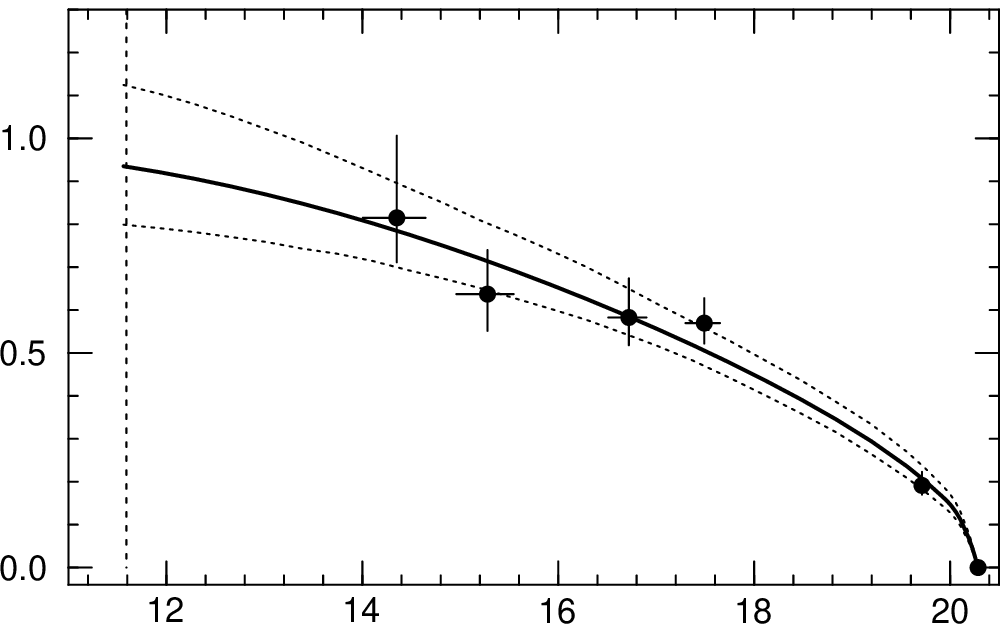}
\point 0 0.56 {\begin{sideways}
$d\Gamma/dq^2\ (|V_{ub}|^2 10^{-12}\gev^{-1})$\end{sideways}}
}\hss}
\hbox to\hsize{\hss
\hbox to\unit{\kern0.5\unit$q^2\ (\!\gev^2)$\hfill}\hss}
\caption[]{Differential decay rate as a function of $q^2$ for the
semileptonic decay $\bar B^0\to\rho^+l^-\bar\nu_l$, taken
from~\cite{ukqcd:btorho}. Points are measured lattice data, solid
curve is fit from \eqref{distr2} with parameters given in
\eqref{ab-vals}. The dashed curves show the variation from the
statistical errors in the fit parameters. The vertical dotted line
marks the charm endpoint.}
\label{fig:vub}
\end{figure}

From lattice simulations we can obtain the form factors only for part
of the physical phase space.  In order to control discretization
errors we require that the three-momenta of the $B$, $\pi$ and $\rho$
mesons be small in lattice units. This implies that we determine the
form factors at large values of momentum transfer $q^2 =
(p_B-p_{\pi,\rho})^2$. Experiments can already reconstruct exclusive
semileptonic $b\to u$ decays (see, for example, the review
in~\cite{jrp:ichep96}) and as techniques improve and new facilities
begin operation, we can expect to be able to compare the lattice form
factor calculations directly with experimental data at large $q^2$. A
proposal in this direction was made by UKQCD~\cite{ukqcd:btorho} for
$\btorho$ decays. To get some idea of the precision that might be
reached, they parametrize the differential decay rate distribution
near $\qsqmax$ by:
\begin{equation}
\frac{d\Gamma(\btorho)}{dq^2}
 =  10^{-12}\,\frac{G_F^2|V_{ub}|^2}{192\pi^3M_B^3}\,
q^2 \, \lambda^{\frac{1}{2}}(q^2)
 \, a^2\left( 1 + b(q^2{-}\qsqmax)\right),
\label{eq:distr2}
\end{equation}
where $a$ and $b$ are parameters, and the phase-space factor $\lambda$
is given by $\lambda(q^2) = (m_B^2+m_\rho^2 - q^2)^2 - 4
m_B^2m_\rho^2$. The constant $a$ plays the role of the IW function
evaluated at $\w=1$ for heavy-to-heavy transitions, but in this case
there is no symmetry to determine its value at leading order in the
heavy quark effective theory. UKQCD obtain~\cite{ukqcd:btorho}
\begin{equation}
\begin{array}{rcl}
a &=& 4.6 \err{0.4}{0.3} \pm 0.6 \gev, \\
b &=& (-8 \err46) \times 10^{-2} \gev^2.
\end{array}
\label{eq:ab-vals}
\end{equation}
The fits are less sensitive to $b$, so it is less well-determined. The
result for $a$ incorporates a systematic error dominated by the
uncertainty ascribed to discretization errors and would lead to an
extraction of $|V_{ub}|$ with less than 10\% statistical error and
about 12\% systematic error from the theoretical input.  The
prediction for the $d\Gamma/dq^2$ distribution based on these numbers
is presented in \figref{vub}.  With sufficient experimental data an
accurate lattice result at a single value of $q^2$ would be sufficient
to fix $|V_{ub}|$.

In principle, a similar analysis could be applied to the decay
$\btopi$. However, UKQCD find that the difficulty  of performing the
chiral extrapolation to a realistically light pion from the unphysical
pions used in the simulations makes the results less certain. 
The $B\to\pi$ decay also has a smaller fraction of events at high
$q^2$, so it will be more difficult experimentally to extract
sufficient data in this region for a detailed comparison.

We would also like to know the full $q^2$ dependence of the form
factors, which involves a large extrapolation in $q^2$ from the high
values where lattice calculations produce results, down to $q^2=0$. In
particular the radiative decay $\btokstargamma$
occurs at $q^2=0$, so that existing lattice simulations cannot
make a direct calculation of the necessary form factors.

\begin{figure}
\hbox to\hsize{\hss\vbox{\offinterlineskip
\unit=0.6\hsize
\epsfxsize=\unit\epsffile{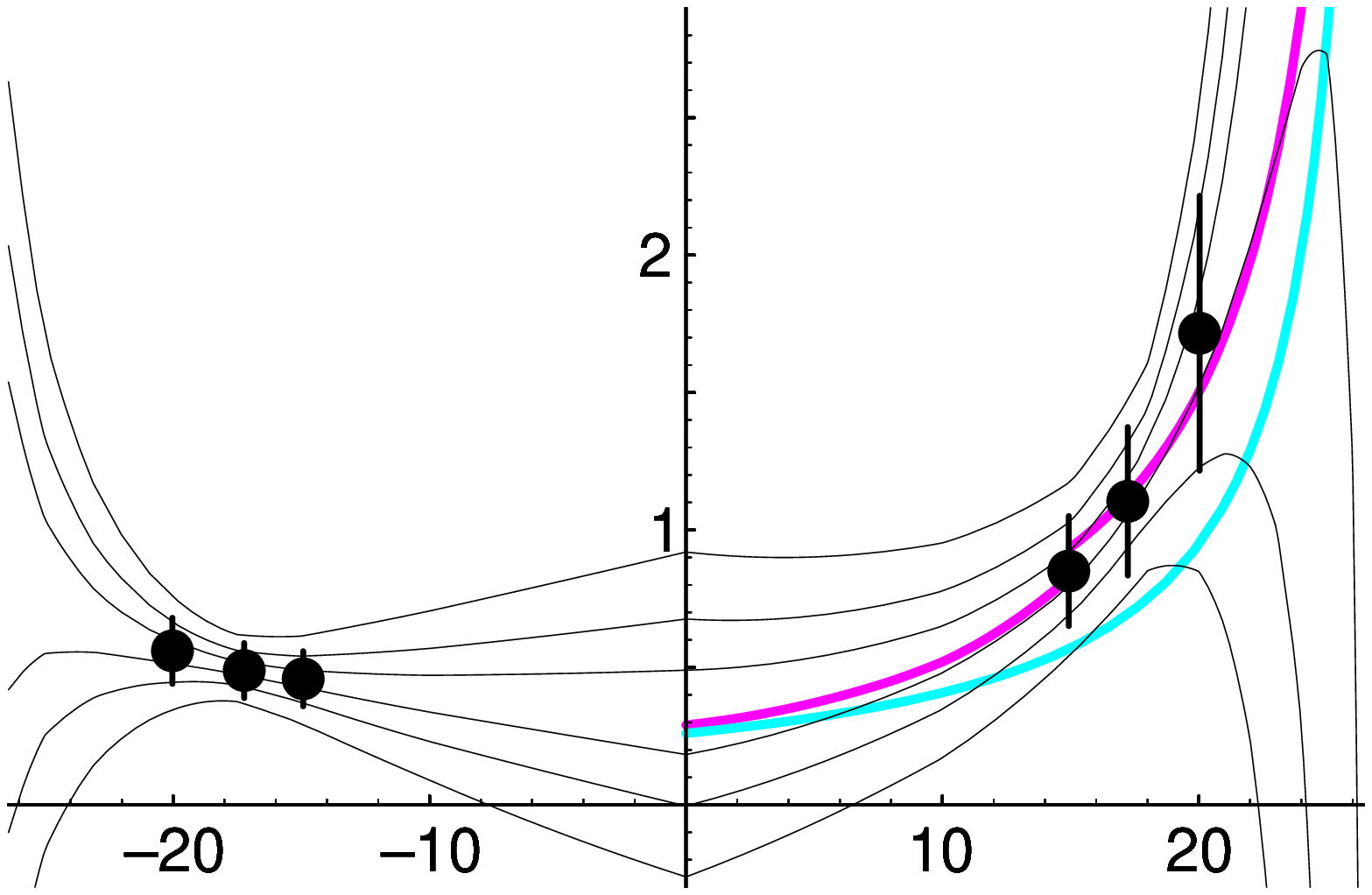}
\point 0.55 0.74 {f^+(q^2)}
\point 0.3 0.74 {f^0(|q^2|)}
\point 0.82 0.07 {q^2\ (\!\gev^2)}
}\hss}
\caption[]{Bounds on $f^+$ and $f^0$ for $\btopi$ from dispersive
constraints~\cite{lpl:bounds}. The data points are from
UKQCD~\cite{ukqcd:hlff}, with added systematic errors. The pairs of
fine curves are, outermost to innermost, 95\%, 70\% and 30\% bounds.
The upper and lower shaded curves are
light-cone~\protect\cite{bbkr:B-Bstar-pi-couplings} and
three-point~\protect\cite{PaB93} sum rule results respectively.}
\label{fig:btopi-bounds}
\end{figure}

An interesting approach to this extrapolation problem has been applied
by Lellouch~\cite{lpl:bounds} for $\btopi$. Using dispersion relations
constrained by UKQCD lattice results at large values of $q^2$ and
kinematical constraints at $q^2=0$, one can tighten the bounds on form
factors at all values of $q^2$. This technique relies on perturbative
QCD in evaluating one side of the dispersion relations, together with
general properties of field theory, such as unitarity, analyticity and
crossing. It provides model-independent results which are illustrated
in \figref{btopi-bounds}. The results (at 50\% CL ---
see Ref.~\cite{lpl:bounds} for details) are
\begin{eqnarray}
f^+(0) &=& 0.10\hbox{--}0.57, \\
\Gamma(\btopi) &=& 4.4\hbox{--}13\, |V_{ub}^2| \, \mathrm{ps}^{-1}.
\end{eqnarray}
Unfortunately these bounds are not very restrictive when constrained
by existing lattice data. In principle, this method can be applied to
$B\to\rho$ decays also, but is more complicated there, and the
calculations have yet to be performed. Recently,
Becirevic~\cite{becirevic:bksbounds} has applied the method for
$\btokstargamma$, using APE lattice results as constraints. However,
he has not applied the kinematic constraint from
\eqref{ffkinconstraints} and the resulting bounds are not informative:
they become so, however, once he uses a light-cone sum rule evaluation of
$T_1(0) = iT_2(0)$ as an additional constraint. These dispersive
methods can be used with other approaches in addition to lattice
results and sum rules, such as quark models, or even in direct
comparisons with experimental data, to check for compatibility with
QCD and to extend the range of results.

For now we must rely on model input to guide $q^2$ extrapolations.  We
can ensure that any assumed $q^2$-dependence of the form factors is
consistent with the requirements imposed by heavy quark symmetry, as
shown in \eqref{scaling}, together with the kinematical relations of
\eqref{ffkinconstraints}. Even with these constraints, however,
current lattice data do not by themselves distinguish a preferred
$q^2$-dependence. Fortunately, more guidance is available from
light-cone sum rule analyses~\cite{abs,patricia} which lead to scaling
laws for the form factors at fixed (low) $q^2$ rather than at fixed
$\w$ as in \eqref{scaling}. In particular all form factors scale like
$M^{-3/2}$ at $q^2=0$:
\begin{equation} f(0) \Theta = M^{-3/2} \gamma_f \left( 1
+ {\delta_f\over M} +
 {\epsilon_f\over M^2} + \cdots \right).
\label{eq:scalingm}
\end{equation}
It is therefore important to use ans\"atze for the form factors
compatible with as many of the known constraints as possible.

\begin{table}
\vspace{-8pt}
\caption[]{Lattice results for $\btopi$ using various ans\"atze for
the form factor $f^+$. The decay rates are values for
$\Gamma(\btopi)/\vub^2 {\rm ps}^{-1}$. ELC~\cite{elc:hl-semilept} and
APE~\cite{ape:hl-semilept} results are from their method `b', which
uses the heavy quark scaling laws of \eqref{scaling} to extrapolate
from $D$- to $B$-mesons at fixed~$\w$.}
\label{tab:btopi-results}
\kern1em
\begin{center}
\begin{tabular}{rlllllll} \hline
\ts  & Yr & $\beta$ & $\csw$ & norm & Rate & $f^+(0)$ & \\[0.4ex]
\hline
\ts UKQCD~\cite{ukqcd:hlfits}
 & 97 & 6.2 & 1 & rel & $8.5\errp{33}{14}$ & 0.27(11) & \\
%\multicolumn{4}{l}{dipole/pole fit to $f^+$/$f^0$}\\[1ex]
WUP~\cite{wup:hl-semilept}
 & 97 & 6.3 & 0 & nr & & $0.43(19)$ \\
APE~\cite{ape:hl-semilept}
 & 95 & 6.0 & 1 & rel & $8\pm 4$ & 0.35(8) & \\
%\multicolumn{4}{l}{$q^2{\simeq} 20.4\gev^2$, pole fit,
%$m_\mathrm{p}{=}5.32(1)\gev$}\\[1ex]
ELC~\cite{elc:hl-semilept}
 & 94 & 6.4 & 0 & rel & $9\pm 6$ & 0.30(14)(5) & \\[0.4ex]
\hline
\end{tabular}
\end{center}
\end{table}

Lattice results for $\btopi$, $\btorho$ and $\btokstargamma$ are
reported in Tables~\ref{tab:btopi-results}, \ref{tab:btorho-results}
and~\ref{tab:btokstargamma} respectively. ELC~\cite{elc:hl-semilept}
and APE~\cite{ape:hl-semilept} fit lattice data for the semileptonic
decays at a single value of $q^2$ to a simple pole form with the
appropriate pole mass also determined by their data. For the $f^0$ and
$A_1$ form factors, this is consistent with heavy quark symmetry
requirements, kinematic relations and light-cone scaling relations at
$q^2=0$, \eqref{scalingm}, but for the other form factors it is not
simultaneously consistent. The WUP~\cite{wup:hl-semilept} results are
found by scaling form factors at $q^2=0$ from results with quark
masses around the charm mass to the $b$-quark mass. However, the
scaling laws used do not follow the light-cone scaling relations of
\eqref{scalingm}.

\begin{table}
\vspace{-8pt}
\caption[]{$\btorho$ results from lattice simulations. The decay rates
are values for $\Gamma(\btorho)/|V_{ub}|^2
\mathrm{ps}^{-1}$. ELC~\cite{elc:hl-semilept} and
APE~\cite{ape:hl-semilept} results are from their method `b', which
uses the heavy quark scaling laws of \eqref{scaling} to extrapolate
from $D$- to $B$-mesons at fixed~$\w$.}
\label{tab:btorho-results}
\kern1em
\begin{center}
\begin{tabular}{rllllllll} \hline
\ts & Yr & $\beta$ & $\csw$ & norm & Rate & $V(0)$ & $A_1(0)$ &
 $A_2(0)$ \\[0.4ex]
\hline
\ts UKQCD~\cite{ukqcd:hlfits} & 97 & 6.2 & 1 & rel &
 $16.5\errp{35}{23}$ & 0.35\errp65 & 0.27\errp54 & 0.26\errp53 \\
%\multicolumn{5}{l}{combined fit to $P\to V$ ff's}\\
WUP~\cite{wup:hl-semilept} & 97 & 6.3 & 0 & nr &
  & 0.65(15) & 0.28(3) & 0.46(23) \\
APE~\cite{ape:hl-semilept} & 95 & 6.0 & 1 & rel &
 $12\pm 6$ & 0.53(31) & 0.24(12) & 0.27(80) \\
%\multicolumn{5}{l}{pole fits to $0\to1$ channel at a single $q^2$}\\
ELC~\cite{elc:hl-semilept} & 94 & 6.4 & 0 & rel &
 $14\pm12$ & 0.37(11) & 0.22(5) & 0.49(21)(5) \\[0.4ex]
%\multicolumn{5}{l}{pole fits to $0\to1$ channel at a single $q^2$}\\
\hline
\end{tabular}
\end{center}
\end{table}

\begin{figure}
\unit=0.7\hsize
\hbox to\hsize{\hss
\vbox{\offinterlineskip
\epsfxsize=\unit\epsffile[-20 50 288 236]{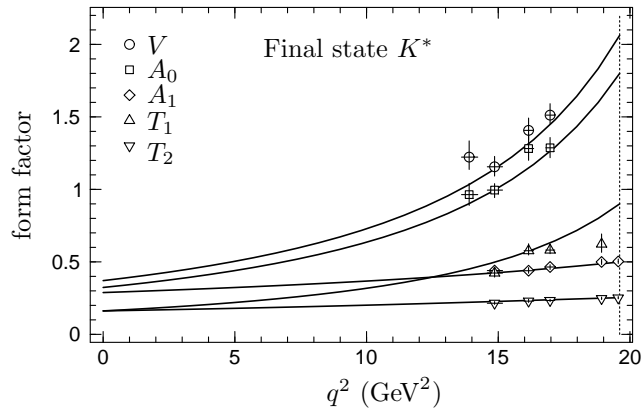}
\point 0.4 0.54 {{\rm Final\ state\ }K^*}
\point 0.215 0.543 V
\point 0.215 0.505 {A_0}
\point 0.215 0.465 {A_1}
\point 0.215 0.422 {T_1}
\point 0.215 0.374 {T_2}
\point 0 0.425 {\begin{sideways}{form factor}\end{sideways}}
%\kern0.5em
\hbox to\unit{\kern0.5\unit$q^2\ (\!\gev^2)$\hfill}
}\hss}
\caption[]{UKQCD~\cite{ukqcd:hlfits} fit to the lattice predictions
for $A_0$, $A_1$, $V$, $T_1$ and $T_2$ for a $K^*$ meson final state
assuming a pole form for $A_1$. $A_2$ is not reliably extracted from
the lattice data so is not used in the fit. The dashed vertical line
indicates $\qsqmax$.
\label{fig:ukqcd-kstarfit}}
\end{figure}

The latest UKQCD study~\cite{ukqcd:hlfits} uses models consistent with
all constraints, including the light-cone sum rule scaling
relations~\footnote{In the extrapolation of the raw lattice results to
the $B$-meson mass, this analysis also applies the condition in
\eqref{a1t2vt1hqs}, that the pairs $(V,2T_1)$ and $(A_1,2iT_2)$ should
agree at fixed~$\w$ in the infinite heavy mass limit.}. For $\btopi$
UKQCD use a dipole(pole) model for $f^+$($f^0$), while for $\btorho$
and $\btokstargamma$ (collectively denoted by $B\to V$) they use a
model inspired by Stech~\cite{stech} which expresses \emph{all} the
form factors for the semileptonic and radiative decays of a heavy
pseudoscalar meson to a light vector meson in terms of a single
function, taken to be a simple pole form for $A_1$. In lattice
calculations one has the freedom to adjust hadron masses by tuning the
quark masses used in the simulation. UKQCD use this freedom to perform
a combined fit within their model for all the $B\to V$ form factors
simultaneously, first with $V=\rho$ and then with $V=K^*$. From the
first fit they obtain the form factors for $\btorho$ and from the
second they obtain results for $\btokstargamma$. The combined fit in
the $K^*$ case is illustrated in \figref{ukqcd-kstarfit}. The figure
demonstrates the large extrapolation needed to reach $q^2=0$.

Our preferred results for $\btopi$ and $\btorho$ come from the UKQCD
constrained fits~\cite{ukqcd:hlfits}. Their values for the form
factors extrapolated to $q^2=0$ agree well with light-cone sum rule
calculations, which work best at low $q^2$. The fitted form factors
also agree with experimental results for the rates and ratio-of-rates
of these semileptonic decays. However, we emphasise that the
extrapolated form factors are no longer model independent. 

There are also preliminary results for heavy-to-light form
factors from FNAL, JLQCD and a Hiroshima-KEK group (see the reviews
in~\cite{jmfstlouis,onogi:lat97}) and the different lattice
calculations are in agreement for the form factors at large
$q^2$ where they are measured.

Table~\ref{tab:btokstargamma} lists the values of $T(0)\equiv T_1(0) =
iT_2(0)$ for $\btokstargamma$, together with the directly measured
$T_2(\qsqmax)$. All groups find that $T_2$ has much less $q^2$
dependence than $T_1$, although the lattice data again do not
themselves distinguish a preferred $q^2$ dependence. In order to make
a distinction, one can apply the light-cone sum rule scaling relation
at $q^2=0$, see \eqref{scalingm}, which states that $T(0)$ has a
leading $M^{-3/2}$ behaviour. In the table we list results from form
factor fits which satisfy this scaling law. Our preference is to quote
the UKQCD~\cite{ukqcd:hlfits} result (with statistical error only)
from the combined fit to $B\to V$ decays described above:
\begin{equation}
T(0) = 0.16\errp21.
\label{eq:t0best}
\end{equation}
Using this value to evaluate the ratio (given at leading order in QCD
and up to $O(1/m_b^2)$ corrections~\cite{ciuchini94})
\begin{equation}
R_{K^*} = {\Gamma(\btokstargamma)\over\Gamma(b\to s\gamma)} =
  4\left(\frac{m_B}{m_b}\right)^3
  \left(1-\frac{m^2_{K^*}}{m^2_B}\right)^3 |T(0)|^2
\end{equation}
results in
\begin{equation}
R_{K^*} = 16\errp43 \%,
\end{equation}
which is consistent with the experimental result $18(7)\%$ from
CLEO~\cite{cleo:rkstar}. Discrepancies between $R_{K^*}$ calculated
using $T(0)$ and the experimental ratio
$\Gamma(\btokstargamma)/\Gamma(b\to s\gamma)$ could reveal the
existence of long-distance effects in the exclusive decay. It has been
proposed that these effects may be significant for the process
$\btokstargamma$~\cite{gp:longdist1}$^-$\cite{abs:longdist}, but
within the precision of the experimental and lattice results, there is
no evidence for them.

\begin{table}
\vspace{-8pt}
\caption[]{Lattice results for $\btokstargamma$. Values for
$T(0)\equiv T_1(0) = iT_2(0)$ are quoted only from models which
satisfy the light-cone sum rule scaling relation, \eqref{scalingm}, at
$q^2=0$.}
\label{tab:btokstargamma}
\kern1em
\begin{center}
\begin{tabular}{rllllll}\hline
\ts  & Yr & $\beta$ & $\csw$ & norm & $T(0)$ &
 $T_2(\qsqmax)$ \\[0.4ex]
\hline
\ts UKQCD~\cite{ukqcd:hlfits} & 97 & 6.2 & 1 & rel &
 0.16\errp21 & 0.25(2) \\
LANL~\cite{lanl:wme-lat95} & 96 & 6.0 & 0 & nr &
 0.09(1) \\
APE~\cite{ape:bsg-clover} & 96 & 6.0 & 1 & rel &
 0.09(1)(1) \\
BHS~\cite{bhs:bsg} & 94 & 6.0 & 0 & nr &
 0.101(10)(28) & 0.325(33)(65) \\[0.4ex]
\hline
\end{tabular}
\end{center}
\end{table}

\section{The Parameters of the HQET}
\label{sec:hqet}

The Heavy Quark Effective Theory (HQET) is proving to be a
particularly useful tool for phenomenological studies in charm and
beauty physics, for reviews and references to the original literature
see~\cite{neubert,neubertreview,iw}. In this approach, physical
quantities are calculated as series in inverse powers of the mass(es)
of the heavy quark(s). The non-perturbative strong interaction effects
can be parametrized in terms of matrix elements of local operators,
which appear as factors in the expansion coefficients.  Lattice
simulations of the HQET provide the opportunity of computing these
matrix elements numerically and in this section we briefly describe
some of these calculations. We start with a general discussion 
(for a more detailed presentation and references to the original
literature see Ref.~\cite{ht}) and then we
will illustrate the main points by some important examples.

Consider some physical quantity $\mathcal{P}$ which, using the
operator product expansion can be written as a series in inverse
powers of the mass of the heavy quark, $m_Q$,~\footnote{For simplicity
we assume that $\mathcal{P}$ depends only on the mass of one heavy
quark.}
\begin{equation}
\mathcal{P} = C_1(m_Q^2/\mu^2) \< f|O_1(\mu) |i\> +
 \frac{C_2(m_Q^2/\mu^2)}{m_Q^n}\, \< f|O_2(\mu)|i\> + O(1/m_Q^{n+p})\ ,
\label{eq:ope}
\end{equation}
where $n\ge 1$ and the coefficient functions $C_i$ are independent of
the states $|i\>$ and $|f\>$. The operators $O_{1,2}$ are composite
operators of static heavy quark, light quark and gluon fields,
renormalized at a scale $\mu$. For clarity of notation, we suppress
the dependence of the coefficient functions on the coupling constant,
$\alpha_s(m_Q^2)$.  We assume here that there is only one operator in
each of the first two orders of the expansion. If this is not the
case, then there is an additional mixing of operators, which requires
only a minor modification of the discussion below. We will therefore
not consider this possibility further. The final term of
$O(1/m_Q^{n+p})$ in \eqref{ope} represents the contributions of
operators of even higher dimension which will not be discussed here.

Throughout this discussion we assume that we wish to evaluate the
contributions of $O(\,(\lqcd/m_Q)^n)$ to $\mathcal{P}$, or at least to
improve the precision given by the lowest order contribution.

In lattice simulations we work directly with bare operators
(corresponding to the lattice regularization), so that the
renormalization scale $\mu$ in \eqref{ope} should be replaced by the
cut-off $a^{-1}$.  With a hard ultra-violet cut-off, such as the
lattice spacing, the higher dimensional operators $O_{2,3,\cdots}$ can
mix with lower dimensional ones, and with $O_1$ in particular. By
dimensional arguments it can readily be seen that the mixing
coefficients will diverge as inverse powers of the lattice spacing.
This implies, for example, that
\be
\<\, f\,|\,O_2(a)\,|\,i\,\> = O(a^{-n})\ ,
\label{eq:o2scaling}\ee
and so this matrix element, computed in lattice simulations, is
clearly not a physical quantity. These power divergences are
subtracted in the perturbative matching procedure in which the Wilson
coefficient functions are computed; specifically the coefficient
function $C_1$ contains terms of $O\big((ma)^{-n}\big)$, which cancel
the power divergences present in $\bra f O_2\ket i$. The perturbative
series for $C_1$ is evaluated only at some low order (typically one
loop). Since $(1/m_Qa)^n$ is much larger than $(\lqcd/m_Q)^n$, which
is the order of the terms we are trying to evaluate, the truncation of
the perturbative series for $C_1$ leads to a significant loss of
precision~\footnote{The connection of the power divergences and
renormalons is explained in
Refs.~\cite{ht}$^-$\cite{melbourne}.}. This is the principal problem,
in attempts to quantify power corrections to hard scattering and decay
amplitudes in general.  We mention in passing that although the
discussion here is in the context of lattice calculations, analogous
problems occur also in the continuum regularization schemes, such as
dimensional regularization~\cite{ht}.

In some cases the operator $O_2$ is protected from mixing under
renormalization with $O_1$, because of the presence of some symmetry.
An important example of this in heavy quark physics is the
chromomagnetic operator $\bar h\sigma_{ij}G^{ij}h$ (where $h$
is the field of a static quark and $G^{\mu\nu}$ is the gluon field
strength tensor), which cannot mix with the lower dimensional operator
$\bar h h$ because it has a different spin structure. In such cases
the corresponding problem of power divergences (and renormalon
singularities) does not arise. For the remainder of the discussion we
assume that $O_2$ and $O_1$ have the same quantum numbers, so that
they can mix under renormalization.  Another important exception to
the general discussion is the difference of matrix elements of the
kinetic energy operator taken between different hadronic states. In
this case the higher dimensional operator, $\bar h \vec D^2h$, can mix
with the lower dimensional one, $\bar hh$, but as the latter is a
conserved current it has the same matrix element between all single
hadron states. Thus the corresponding power divergences cancel in the
difference of matrix elements. In all of these exceptions the matrix
elements of the higher dimensional operators (or linear combinations
of matrix elements) are also the leading contribution to some physical
quantity (for example, the chromomagnetic operator gives the leading
contribution to the $B^*$--$B$ mass splitting).

In the following subsections we consider 3 parameters which occur
frequently in the evaluation of physical quantities using the HQET:
\begin{itemize}
\item $\overline{\Lambda}$, the binding energy of a heavy quark 
in a heavy hadron; 
\item $\lambda_1$, the kinetic energy of the heavy quark;
\item $\lambda_2$, the matrix element of the chromomagnetic operator.
\end{itemize}
The binding energy $\overline{\Lambda}$ is defined as the  difference
of the masses of the heavy hadron and the heavy quark. In practice, the
heavy quark mass must be defined at short distances (otherwise
experimentally measurable quantities cannot be expressed in terms of
the mass using perturbation theory), and below we will use the $\MSbar$
mass. The usefulness of introducing $\bar\Lambda$ is then unclear, and
we present the discussion simply in terms of the mass itself. In all
the three examples to be discussed  below, results are presented from
lattice calculations using static heavy quarks.

\subsection{The Evaluation of the Mass of a Heavy Quark}
\label{subsec:mb}

The first example which we consider is the determination of the mass
of a heavy quark (e.g. the $b$-quark), up to, and including the 
terms of $O(\Lambda_{\mathrm{QCD}})$, but neglecting terms of
$O(\Lambda_{\mathrm{QCD}}^2/m_b)$~\footnote{A formulation of this
problem in terms of an explicit operator product expansion can be 
found in Refs.~\cite{beneke,ms}.}. 

In section~\ref{sec:fb}, the computation of the decay constant of 
a meson containing a static heavy quark was discussed. This parameter 
is obtained by evaluating correlation functions of the form:
\be
C(t) = \sum_{\vec x} \langle 0\,|\,A_4(\vec x, t)\, 
A_4(\vec 0,0)\,|\,0\rangle\ 
\label{eq:ct}\ee
in lattice simulations of the HQET, with Lagrangian density
\be
{\cal L} = \bar h D_4 h\ ,
\label{eq:lhqet}\ee 
where $h$ represents the field of the static quark.
For sufficiently large values of the time $t$, 
\be
C(t) \simeq Z^2\ e^{-{\cal E}\,t} + \cdots\ ,
\label{eq:ctasymp}\ee
where the ellipsis represents contributions from excited states.
The value of $f_B$ in the static limit is obtained from the 
prefactor $Z$. In addition, however, from the exponent ${\cal E}$ 
it is possible to obtain the $O(\Lambda_{\mathrm{QCD}})$
contribution to $m_b$. Performing the matching of the heavy quark propagator 
in full QCD and in the HQET gives~\cite{cgms}
\be
{\cal E} = m_B - (m_b^{{\rm pole}} - \delta m)\ ,
\label{eq:cale}\ee
where $m_B$ is the mass of the $B$-meson 
and $m_b^{{\rm pole}}$ is the pole mass of
the $b$-quark. Although we have chosen the bare Langrangian
(\ref{eq:lhqet}) to have no mass term, higher order perturbative
corrections generate such a term and 
\begin{equation} 
\delta m =
\sum_{n=1}^{\infty}\left(\frac{\alpha_s}{4\pi}\right)^n\,\frac{X_n}{a}
\label{eq:deltam}
\end{equation}
represents the perturbation series generating the mass. Both ${\cal
E}$ and $\delta m$ diverge linearly with the lattice spacing, so that
they are not physical quantities. $m_b^{\rm{pole}}$ is also
unphysical, and in perturbation theory contains renormalon
ambiguities~\cite{beneke,bsuv}. These ambiguities cancel those also
present in $\delta m$ in the difference on the right-hand side of
Eq.~(\ref{eq:cale}).
Thus it is possible to determine the value
of a physical (short-distance) definition of the quark mass, such
as $\overline{m}_b\equiv m_b^{\overline{MS}}(m_b^{\overline{MS}})$
from the computed value of ${\cal E}$. 
In practice, however, the subtraction of the linear divergence,
which is performed in perturbation theory leads to large numerical
cancellations and hence to significant uncertainties. 
In reference~\cite{gms1} it was found that
\be
\overline{m}_b= 4.15\pm 0.05\pm 0.20\ {\rm GeV}.
\label{eq:mbvalue}\ee 
The first error on the right-hand side of Eq.~(\ref{eq:mbvalue}) is
due to uncertainties in the lattice evaluation of ${\cal E}$ and in
the value of the lattice spacing. However, the larger error of $200\mev$
or so is the estimate of the uncertainty due to the truncation of the 
perturbation series for $\delta m$ at one-loop order. 
Evaluation of higher order terms in this series, perhaps using
the methods developed in Ref.~\cite{langevin}, based on the Langevin
stochastic formulation of lattice QCD, is urgently needed to reduce
the uncertainty in the computed value of $\overline{m}_b$.

\subsection{Kinetic Energy of a Heavy Quark}
\label{subsec:lambda1}

The next example which we will consider is the evaluation of the
kinetic energy of the heavy quark $\lambda_1$, which appears in many
applications of the HQET~\cite{neubert}. On the lattice we start with
the evaluation of the matrix element of the bare operator
\beqn \lambda_1^{\mathrm{bare}} & = & 
- \,\frac{\langle B| \bar h (i\vec D)^2 h|B\rangle}{2
M_B}\label{eq:l1baredef}\\  
& = & -(0.69\pm 0.03\pm 0.03)\, a^{-2}
\label{eq:l1bare}\eeqn 
where the numerical result is taken from a simulation on a $24^3\times
60$ lattice at $\beta = 6.0$ with the SW-action~\cite{gms2}. 
Taking the lattice spacing to be $2.0\pm 0.2$~GeV, we see that the
magnitude of the result is about 2.8~GeV$^2$, to be compared to the
expected physical corrections of $O(\lqcd^2)\sim$ 0.1 GeV$^2$. Of
course, the large result is due to the presence of power divergences,
in this case they are quadratic, i.e.\ they are of $O(a^{-2})$. In
one-loop perturbation theory, the power divergence is equal to
$-5.19\,\alpha_s\,a^{-2}$, which, depending on the value taken for the
coupling constant $\alpha_s$, is in the range $(0.67-0.93)a^{-2}$ (the
choice of a suitable definition of the coupling constant is a
representation of our ignorance of the higher order perturbative
corrections, the range given here comes from frequently used
definitions).  The uncertainty is greater than the terms we are trying
to evaluate which are of $O(\Lambda_{\mathrm{QCD}}^2)$.  Clearly, in
order for the lattice result to be useful for phenomenological
applications, the perturbative calculations must be performed to
higher orders, which is the main conclusion of this subsection.

It is also possible to subtract the power divergences non-perturbatively. 
In Refs.~\cite{ms,cgms,gms2} a subtracted kinetic energy operator,
\begin{equation}
\bar h \vec D_S^2 h\equiv 
\bar h\vec D^2h  - \frac{c}{a^2}\bar hh\ , 
\label{eq:d2sub}\end{equation} 
was defined, with the subtraction contant $c$ fixed by imposing that
the matrix element of this operator vanishes between quark states at
rest (in the Landau Gauge). The corresponding value of $\lambda_1$,
which is now free of quadratic divergences, was found to
be~\footnote{Note that the central value in Eq.~(\ref{eq:l1result})
has the opposite sign to that of many other estimates using various
definitions of $\lambda_1$.}
\begin{eqnarray}
\lambda _1 & = & a^{-2}Z_{\vec D^2_S}(a^2\lambda_1^{\mathrm{bare}}
- a^2 c)\label{eq:l1}\\ 
& = & 0.09\pm 0.14\ \mathrm{GeV^2}\ .
\label{eq:l1result}\end{eqnarray}
Of course the large relative error in Eq.~(\ref{eq:l1result}) is due
to the large cancelation between the two terms in the parentheses in
Eq.~(\ref{eq:l1}). In Eq.(\ref{eq:l1}), $Z_{\vec D^2_S}$ is the
normalization constant required to obtain the continuum, $\MSbar$,
value of $\lambda_1$ from the subtracted lattice one. 

The difficulties described here arise because of the mixing of the 
kinetic energy operator $\bar h(i\vec D)^2 h$ with $\bar hh/a^2$.
Since $\bar hh$ is a conserved current in the HQET, its matrix elements
are the same between all hadronic states. This means that the difference 
of the matrix elements of the kinetic energy operator between any two 
different beauty hadrons is a physical quantity, and 
for example in Ref.~\cite{gms2} it was found that 
\be
\lambda_1(B_s)-\lambda_1(B_d) = -0.09\pm 0.04\ \mathrm{GeV^2}\ .
\label{eq:l1bsbd}\ee
This difference is the leading contribution to the following
combination of masses
\begin{equation}
\lambda_1(B_s)-\lambda_1(B_d)  = 
\frac{\overline{m}_{B_s} - \overline{m}_{B_d}
- (\overline{m}_{D_s} - \overline{m}_{D_d})}{\frac{1}{2}
\left(\frac{1}{m_D}-\frac{1}{m_B}\right)}
+ O\left(\frac{\lqcd^3}{m_Q}\right)\ ,
\label{eq:l1bsbdth}\end{equation}
where, for example, $\overline{m}_B$ is the spin averaged mass
of the $B$-meson (with the corresponding light valence quark)
\begin{equation}
\overline{m}_B = \frac{1}{4}(3m_{B^*} + m_B)\ ,
\label{eq:spinavdef}\end{equation}
and $m_Q$ is the mass of the heavy quark $Q$
($Q=b$ or $c$). Where appropriate, we have included the subscript $d$
or $s$ to denote the presence of the corresponding light valence
quark.  The experimental value of the first term on the right hand
side of Eq.~(\ref{eq:l1bsbdth}) is $-0.06\pm0.02$\,GeV$^2$, in very
good agreement with the result in Eq.~(\ref{eq:l1bsbd}). It must
however be remembered that the $O(\lqcd^3/m_c)$ corrections to this
result may be significant.

\subsection{$\lambda_2$; The Matrix Element of the Chromomagnetic
Operator}

The chromomagnetic operator $\bar h \frac{1}{2}\sigma_{ij}G^{ij} h$,
does not mix with the operator $\bar hh$, and hence its matrix elements
are free of power divergences. The parameter $\lambda_2$, defined
as
\begin{equation}
\lambda_2\equiv\frac{1}{2m_B}\,
\frac{1}{3} \langle B\,|\bar h\frac{1}{2}\sigma_{ij}G^{ij} h
|\,B\rangle
\label{eq:lambda2def}\end{equation}
gives the first term in the hyperfine splitting in the $B$-meson system:
\begin{equation}
m_{B^*}^2-m_B^2 = 4\lambda_2\ .
\label{eq:hfs}\end{equation}
The lattice results for $\lambda_2$ are all significantly smaller (by
almost a factor of two) than the values deduced from the
physical masses of the $B^*$ and $B$ mesons. In two recent lattice 
computations the authors found:
\begin{equation}
m_{B^*}^2-m_B^2 =
 \cases{0.28 \pm 0.02\pm 0.04\gev^2
       &Ref.~\cite{ukqcd:staticb}\cr
        0.28 \pm 0.06\gev^2
       &Ref.~\cite{gms2}\cr}
\end{equation}
to be compared to the experimental value of $0.485\pm 0.005\gev^2$,
see Ref.~\cite{pdg}.

One possible source for the discrepency between the lattice results
and the experimental value is the unusually large one-loop
contribution to the renormalization constant relating the lattice and
continuum chromomagnetic operator~\cite{fh}. This renormalization
constant is about 1.85 at one-loop order, and so one may wonder
whether the higher order terms might give a significant contribution.
Other possible contributions to the discrepency might be the use of
the quenched approximation, or that the relation between $\lambda_2$
and the hyperfine splitting may be significantly modified by higher
order corrections in $1/m_b$. It is important to clarify the source of
this discrepency.

Lattice calculations of the hyperfine splitting using propagating
heavy quarks also give a result which is smaller than the experimental
one. This is a different problem, however, which is related to the
presence of a spurious chromomagnetic term of $O(a)$ present in the 
lattice action. This interpretation is confirmed by the fact that
the computed value of the splitting increases as the action is ``improved''
in agreement with expectations~\cite{hfs,km}.

\section{Exclusive Non-Leptonic Decays of Heavy Mesons}
\label{sec:exclusive}

Exclusive non-leptonic decays are, in principle, an important source
of fundamental information on the properties of weak decays of heavy
quarks. Unfortunately our current theoretical understanding of the
non-perturbative QCD effects in these processes is rather primitive,
and we are forced to make assumptions based on factorization and/or
quark models~\cite{nst}. Lattice computations of the corresponding
matrix elements are also difficult~\cite{mate}. They need to be
performed in Euclidean space, where there is no distinction between
{\em in-} and {\em out-}states. The quantities which one obtains 
directly in lattice computations are the (real) averages such as
\begin{equation}
\langle\,M_1\,M_2\,|\,{\cal H}_W\,|B\,\rangle
=\frac{1}{2}\big(\,_{\mathrm{in}}
\langle\,M_1\,M_2\,|\,{\cal H}_W\,|B\,\rangle
+\,_{\mathrm{out}}\langle\,M_1\,M_2\,|\,{\cal H}_W\,|B\,\rangle\,\big)\ ,
\label{eq:inout}\end{equation}
where $M_1$ and $M_2$ are mesons. It is therefore not possible to
obtain directly any information about the phase due to final state
interactions, and hence to determine the matrix elements reliably.
Maiani and Testa~\cite{mate} also showed that the  quantities which are
obtained from the large time behaviour of the corresponding 
correlation functions are the unphysical form factors in which the 
final state mesons are at rest, e.g.
\begin{equation}
\langle\,M_1(\vec p_{M_1}=\vec 0)\,M_2(\vec p_{M_2}=\vec 0)
\,|\,{\cal H}_W\,|B\,\rangle\ .
\label{eq:rest}\end{equation}
For $K\to\pi\pi$ decays chiral perturbation theory can then be used to
obtain the physical form factors with reasonable accuracy. For $D$-
and $B$-meson decays this is not possible.

The publication of the Maiani-Testa~\cite{mate} theorem stopped the
exploratory work on the numerical evaluation of two-body
non-leptonic decays.  These early, and not very accurate papers,
studied the non-penguin contributions to $D$-meson decay
amplitudes~\cite{nonlepold}.

The Maiani-Testa theorem implies that it is not possible to obtain the
phase of the final state interactions without some assumptions about
the amplitudes. The importance of developing reliable quantitative
techniques for the evaluation of non-perturbative QCD effects in
non-leptonic decays cannot be overstated, and so attempts to introduce
``reasonable'' assumptions to enable calculations to be performed (and
compared with experimental data) are needed urgently.  Ciuchini et
al.~\cite{ciuchini,silvestrini} have recently shown that by making a
``smoothness'' hypothesis about the decay amplitudes it is possible to
extract information about the phase of two-body non-leptonic
amplitudes. Studies to see whether their proposals are practicable and
consistent are currently beginning.

\section{Conclusions}
\label{sec:concs}

The decays of heavy quarks, and the $b$-quark in particular, provide
a powerful laboratory to test the standard model of particle physics,
to determine its parameters (in particular, the elements of the
CKM-matrix) and to look for signatures of new
physics. Non-perturbative QCD effects in these weak decays are the
major difficulty in interpreting the present and future experimental
data. Lattice computations provide the opportunity to quantify these
effects, and much effort is being devoted to the evaluation of the
decay amplitudes of heavy quarks. In this article we have reviewed the
status of these calculations, presenting an overview of the results
for leptonic decay constants, the amplitudes for $B-\bar B$ mixing,
the form factors for semileptonic and rare radiative decays and the
parameters $\bar\Lambda$, $\lambda_1$ and $\lambda_2$ of the HQET.
These successful calculations are playing a central r\^ole in 
phenomenological studies.

As explained in \secref{exclusive}, relatively little progress has
been achieved so far in understanding exclusive non-leptonic decays on
the lattice, or indeed any physical process in which there are two
hadrons in the initial or final state. It is very important to try to
overcome the theoretical difficulties which have so far prevented this
very important class of processes from being amenable to study in
lattice simulations.

Lattice simulations are a ``first principles'' non-perturbative
technique for the evaluation of the strong interaction effects in weak
decay amplitudes (and in those of heavy quarks in
particular). However, the precision of the results is limited by the
available computing resources, and the priority is to reduce and
control the systematic uncertainties, especially those due to the
quenched approximation. This is discussed in some detail in
\subsecref{uncert} above. During the next few years, studies with
dynamical fermions will continue and improve, enabling us to begin
seriously quantifying the effects of the quenched approximation.  The
very significant progress achieved in recent years in reducing the
errors due to the finiteness of the lattice spacing, makes this task
easier, allowing for meaningful studies on coarser lattices than would
otherwise be the case. Once precise studies with dynamical quarks
become possible, lattice QCD together with large scale numerical
simulations will be a truly and fully quantitative technique for the
evaluation of the effects of the strong interactions, with no model
assumptions or parameters. Although this will still take a few years,
we have tried to demonstrate in this review that lattice simulations,
with their expected $O(10$--$20\%)$ systematic uncertainties for most
physical quantities, are already a most reliable and useful tool.

\section*{Acknowledgements}
We warmly thank B.~Hill, L.P.~Lellouch, G.~Martinelli, J.~Nieves,
H.~Wittig and all our colleagues from the APE, ELC, FNAL and UKQCD
collaborations from whom we have learned so much. We also thank
C.~Bernard, V.~Braun, P.~Drell, S.~Gottlieb, S.~G\"usken,
S.~Hashimoto, H.~Hoeber, P.~Mackenzie, T.~Onogi, A.~Ryd and
R.~Zaliznyak for their help in clarifying important questions.

We acknowledge support from the Particle Physics and Astronomy Research
Council, UK, through grants GR/K55738 and GR/L22744.


\begin{thebibliography}{999}
\bibitem{degrandloft} T.A.~DeGrand and R.D.~Loft, \prd{38} (1988)
  954. Note that the decay constants quoted in this paper have been
  multiplied by $\sqrt2$ to convert to the normalization used here.
\end{thebibliography}
\end{document}